\documentclass[aps,twocolumn,showpacs,amsmath,amssymb,prc]{revtex4-2}
\usepackage{graphicx,here}
\usepackage{enumitem}
\usepackage{multirow}
\usepackage{tikz}
\usepackage{epstopdf}
\usepackage[percent]{overpic}
\usepackage{scrextend}
\usepackage{xcolor,xspace}
\usepackage[ulem=normalem]{changes}
\usepackage{hyperref}
\hypersetup{colorlinks=True,urlcolor=blue,linkcolor=blue,citecolor=blue,filecolor=black}

\colorlet{Changes@Color}{red}
\usepackage{dcolumn,color,footnote,bm,braket}
\usepackage{url,longtable,tabularx}
\usepackage{threeparttable}

\usepackage{fancybox}
\usetikzlibrary{matrix}

\newcommand\+{\dagger}

\begin{document}

\title{Mapped $spdf$ interacting boson model for quadrupole-octupole collective states in nuclei}

\author{K. Nomura}
\email{nomura@sci.hokudai.ac.jp}
\affiliation{Department of Physics, 
Hokkaido University, Sapporo 060-0810, Japan}
\affiliation{Nuclear Reaction Data Center, 
Hokkaido University, Sapporo 060-0810, Japan}

\date{\today}

\begin{abstract}
Dipole bosons are introduced
in the interacting boson model (IBM) by means of the
self-consistent mean-field method.
The constrained mean-field calculations employing a given
nuclear energy density functional yield the potential energy
surfaces in terms of the axially-symmetric quadrupole-octupole,
dipole-quadrupole, and dipole-octupole deformations.
By mapping these energy surfaces onto the
expectation values of the IBM Hamiltonian in the coherent
state of the interacting $s$, $p$, $d$, and $f$ bosons,
strength parameters of the $spdf$-IBM Hamiltonian are determined.
In an illustrative application
to octupole-deformed actinides $^{218-230}$Ra and $^{220-232}$Th,
it is shown that effects of including $p$ bosons in
the IBM mapping are to lower significantly
negative-parity yrast levels, and to improve
descriptions of observed energy-level systematic
in nearly spherical and transitional nuclei, and
of the behaviors of the reduced electric dipole transitions 
and intrinsic dipole moments with neutron number.
\end{abstract}

\maketitle

\section{Introduction}

Low-lying collective states of most nuclei are dominated
by the quadrupole modes including the
anharmonic vibrations of the nuclear surface and
rotational motions.
Higher-order, i.e., octupole, hexadecapole, etc.
deformations also arise and come to play a role in
determining the low-lying nuclear structure.
The octupole collectivity, in particular, corresponds
to a reflection asymmetric (or pear-like) deformation,
which breaks parity.
The octupole modes of collective excitations are
characterized by the appearance of negative-parity
bands that are close in energy to the positive-parity
ground-state bands, forming alternating-parity bands,
and by the appreciable electric octupole ($E3$)
transitions \cite{butler1996,butler2016,butler2020b,butler2024}.

In fact, nuclides that are known to exhibit a static
octupole deformation are rare, the experimental evidence
being suggested in a few radioactive nuclei, e.g.,
$^{220}$Rn \cite{gaffney2013}, $^{224}$Ra \cite{gaffney2013},
$^{144}$Ba \cite{bucher2016}, and $^{146}$Ba \cite{bucher2017}.
These nuclei correspond to the neutron $N$ and/or proton
$Z$ numbers close to 56, 88, 136, at which
nucleon numbers a coupling of two single-particle
orbits that differ by $\Delta j = \Delta l=3\hbar$
occurs, with $j$ and $l$ denoting the quantum numbers
of a single-particle state.
The presence of the octupole correlations also
represents a broader physical significance, as it enhances
the atomic electric dipole moment (EDM),
the observation of which implies violation
of time reversal (T) or charge parity (CP) \cite{engel2025}.
Theoretical investigations for the quadrupole and octupole
deformations and collective excitations have been
made in various nuclear structure models
(see, e.g., Refs.~\cite{butler2016,robledo2019,nomura2020oct,butler2020b,butler2024}
and some related references therein).

The low-lying quadrupole
and octupole states have extensively been studied
by using the framework of
the interacting boson model (IBM) \cite{IBM}.
The IBM comprises the monopole, $s$, and quadrupole, $d$,
bosons, which represent the collective monopole
(with spin and parity $J^{\pi}=0^+$)
and quadrupole ($J^{\pi}=2^+$) pairs of valence nucleons,
respectively \cite{IBM,OAIT,OAI}.
In order to compute negative-parity states
octupole, $f$, bosons, representing the collective octupole
(with $J^\pi=3^-$) nucleon pairs, have been
introduced \cite{IBM,engel1985,engel1987}.
In the last decade, in particular, a method of deriving the
$sdf$-IBM Hamiltonian using the framework of the
nuclear energy density functional (EDF) theory has
been developed \cite{nomura2013oct,nomura2014}.
The first step within this scheme is
the self-consistent mean-field (SCMF) calculation
with constraints on the quadrupole and octupole moments,
which is based on a given nuclear EDF,
e.g., in the nonrelativistic or relativistic regimes.
The self-consistent calculation yields the
potential energy surface (PES) defined in the
quadrupole and octupole deformations.
The next step is to map the PES onto the
corresponding energy surface in the $sdf$-boson system,
by which procedure the strength parameters
of the $sdf$-IBM Hamiltonian are determined.
The method, denoted hereafter as the mapped IBM,
has been applied to study relevance of octupole
correlations in low-energy collective states in sets
of medium-heavy and heavy nuclei
in different mass regions
in which octupole deformations are expected to emerge
(see Refs.~\cite{nomura2023oct,nomura2025rev} for reviews).
This approach, however, has a difficulty of not being
able to reproduce observed electric dipole ($E1$)
transition properties even qualitatively,
and this deficiency has been attributed
to some missing boson degrees of freedom.

Indeed, in the initial implementation of the IBM \cite{engel1985}
in the study of reflection asymmetric nuclear states,
dipole, $p$, bosons, with $J^{\pi}=1^-$,
were considered as a building block in addition to $f$ bosons.
However, while from a microscopic point of view $d$ and $f$ bosons
can be, as mentioned, interpreted to be
collective in nature,
the origin of a $p$ boson appears to have been less obviously
established. That is, it was attributed either to
the spurious center-of-mass motion \cite{engel1987},
or to the giant dipole resonance \cite{sugita1996}.
Mathematically the version of the IBM that includes $p$ bosons
(denoted $spdf$-IBM) furnishes the group SU(16)
\cite{engel1985,engel1987,kusnezov1989,kusnezov1990},
and to produce the rotational limit SU(3) the model
should contain $f$ and $p$ bosons.
Also in realistic applications, the $spdf$-IBM gives improved
descriptions, in particular, of the $E1$
properties of the octupole deformed nuclei over the $sdf$-IBM
\cite{KUSNEZOV1988,sugita1996,zamfir2001,spieker2015,vallejos2021}.
A $p$ or dipole boson degree of freedom can also be
considered to represent an $\alpha$ particle,
and has been introduced to describe molecular-like
structures in light \cite{bijker2017}
and heavy \cite{iachello82,daley1983,spieker2015} nuclei.
In addition, it was shown that the dipole pairs
account for the intrinsic state of
deformed actinides to a large extent, and
should be considered as essential building blocks
as the octupole pairs \cite{otsuka1986,otsuka1988}.
It appears, therefore, that
the IBM model space may need to contain dipole bosons
for the calculations of octupole-deformed nuclei,
and it is of interest to study $p$-boson
effects on various physical observables
of these nuclear systems
in a timely manner.

In the following, dipole ($p$) bosons are
introduced in the aforementioned IBM mapping procedure.
Relevance of including $p$ bosons to reproducing
energy spectra and transition properties
of positive-parity and negative-parity collective states
is addressed by taking as an illustrative example
axially-deformed actinides $^{218-230}$Ra and $^{220-232}$Th.
The PESs as the microscopic input are provided by
the Hartree-Fock-Bogoliubov (HFB) SCMF calculations
\cite{robledo2019} employing the finite range,
Gogny force \cite{Gogny}
with the D1M parametrization \cite{D1M},
performed with constraints on the axial
quadrupole, octupole, and dipole moments.
In the SCMF framework, the dipole deformation
effects on PESs have been
discussed in quite a few instances,
e.g., Refs.~\cite{kowal2012,Jachimowicz2013,dobrowolski2017}.
Also geometry of the $spdf$-IBM Hamiltonian was
studied in the coherent state framework in Ref.~\cite{kuyucak2002}.
Here the quadrupole and octupole
deformations in the fermionic system are associated
with the bosonic counterparts, as has been done
previously \cite{nomura2013oct,nomura2014,nomura2020oct}.
By analogy with these associations,
the bosonic deformation corresponding to
$p$ bosons is assumed to represent
the electric dipole moment in the fermionic system,
which is used as additional
collective coordinate in the SCMF calculations.
It is noted that the mapped $sdf$-IBM calculations
on the above nuclei were made in Ref.~\cite{nomura2020oct}
using the same microscopic input, i.e., Gogny-D1M EDF.
Some of the results reported in that reference are
exploited in the present work, that is,
the quadrupole-octupole PESs,
and derived parameters for the $sdf$-IBM.
To identify $p$-boson effects on various
physical observables, simpler $sdf$-IBM calculations
are also performed, and are compared
with the more realistic
$spdf$-IBM results.

The theoretical procedure is illustrated in Sec.~\ref{sec:model}.
The mapped $spdf$-IBM results for the PESs,
derived parameters, spectroscopic properties
including excitation energies and transition properties,
and detailed energy spectra for a few selected nuclei
are given in Sec.~\ref{sec:results}.
Summary of the main results and concluding remarks
are given in Sec.~\ref{sec:summary}.

\section{Theoretical procedure\label{sec:model}}

The procedure starts with the constrained self-consistent
calculations of the PESs for $^{218-230}$Ra
and $^{220-232}$Th.
In Ref.~\cite{nomura2020oct}
the Gogny-D1M HFB SCMF calculations were made for these nuclei
with the constraints on axially symmetric mass quadrupole $q_{20}$
and octupole $q_{30}$ intrinsic moments,
with the center-of-mass being
fixed at the origin \cite{rayner2012,robledo2013}.
The $q_{20}$ and $q_{30}$ moments are expressed as the
geometrical deformations \cite{BM}
denoted as $\beta_{20}$ and $\beta_{30}$,
respectively:
\begin{eqnarray}
\label{eq:qbeta}
 \beta_{\lambda0}=\frac{\sqrt{(2\lambda+1)\pi}}{3R_0^{\lambda}A}q_{\lambda0}
\end{eqnarray}
with $R_0=1.2A^{1/3}$ fm and $\lambda=2,3$.
The HFB calculations provided PESs in terms of the
$\beta_{20}$ and $\beta_{30}$ deformations.
In the present work, additional Gogny-D1M HFB
calculations are carried out
using the computer program HFBTHO (v4.0)
\cite{hfbtho400} that
include the axial quadrupole, octupole, and dipole deformations
as collective coordinates, and yield two energy surfaces:
one in terms of the dipole and quadrupole deformations,
and the other of the dipole and octupole deformations.
To compute the dipole-quadrupole and dipole-octupole PESs,
the remaining deformation (octupole and quadrupole, respectively)
is fixed at the value corresponding to the global minimum
in the $(\beta_{20},\beta_{30})$-PES obtained in
Ref.~\cite{nomura2020oct}.
In what follows the intrinsic dipole deformation is
conveniently expressed by the symbol $\beta_{10}$,
which is obtained by using the formula of Eq.~\eqref{eq:qbeta}
with $\lambda=1$, as in the case of the quadrupole
and octupole deformations.
Also the $\beta_{\lambda0}$ deformations are
denoted simply as $\beta_{\lambda}$.

The HFB $(\beta_2,\beta_3)$-, $(\beta_1,\beta_2)$-,
and $(\beta_1,\beta_3)$-PESs thus obtained for each nucleus
are associated with the bosonic counterparts in the $spdf$-IBM.
The $spdf$-IBM Hamiltonian adopted in the present study reads
\begin{align}
\label{eq:bh}
 \hat H_{\rm B} = 
&\,
\epsilon_{d} \hat n_{d}
+\epsilon_{f} \hat n_{f} 
+\epsilon_{p} \hat n_{p}
\nonumber\\
&
+ \kappa_{2} \hat Q \cdot \hat Q
  + \kappa_{3} \hat O \cdot \hat O 
+ \kappa_{1} \hat D \cdot \hat D
  + \kappa' \hat L \cdot \hat L \; .
\end{align}
The first, second, and third terms denote the
single $d$, $f$, and $p$ boson number operators,
and are given as $\hat n_d=d^\+\cdot\tilde d$,
$\hat n_f = f^\+ \cdot \tilde f$, and
$\hat n_p = p^\+ \cdot \tilde p$, respectively, 
with $\epsilon_{d}$, $\epsilon_{f}$, $\epsilon_p$
representing the single $d$, $f$, and $p$ boson
energies relative to that of an $s$ boson. 
Note the notations $\tilde d_{\mu} = (-1)^{\mu} d_{-\mu}$,
$\tilde f_{\mu} = (-1)^{3+\mu} f_{-\mu}$, and
$\tilde p_{\mu} = (-1)^{1+\mu} p_{-\mu}$.
The fourth, fifth, and sixth terms
in (\ref{eq:bh}) stand for quadrupole-quadrupole,
octupole-octupole, and dipole-dipole
interactions, respectively,
with the corresponding strength parameters
$\kappa_2$, $\kappa_3$, and $\kappa_1$.
The quadrupole $\hat Q$, octupole $\hat O$, and
dipole $D$ operators read
\begin{align}
\label{eq:q}
\hat Q = 
&\,
s^\+ \tilde d + d^\+ s 
+ \chi_{dd} (d^\+ \times \tilde d)^{(2)}
+ \chi_{pp} (p^\+ \times \tilde p)^{(2)}
\nonumber\\
&
+ \chi_{pf} (p^\+ \times \tilde f + f^\+ \times \tilde p)^{(2)}
+ \chi_{ff} (f^\+ \times \tilde f)^{(2)} \; , 
\end{align}
\begin{align}
\label{eq:o}
\hat O =
&\,
s^\+\tilde f + f^\+ s
+ \chi_{pd}(p^\+\times\tilde d +d^\+\times\tilde p)^{(3)}
\nonumber\\
&
+ \chi_{df}(d^\+\times\tilde f +f^\+\times\tilde d)^{(3)} \; ,
\end{align}
\begin{align}
\label{eq:d}
\hat D =
&\,
s^\+\tilde p + p^\+ s
+ \chi'_{pd}(p^\+\times\tilde d +d^\+\times\tilde p)^{(1)}
\nonumber\\
&
+ \chi'_{df}(d^\+\times\tilde f +f^\+\times\tilde d)^{(1)} \; ,
\end{align}
with $\chi$s being dimensionless parameters.
The last term in Eq.~\eqref{eq:bh}, with $\hat L$ being the boson
angular momentum operator 
\begin{align}
 \hat L=\sqrt{10}(d^\+\times\tilde d)^{(1)}-\sqrt{28}(f^\+\times\tilde f)^{(1)} \; ,
\end{align}
is introduced because it plays an important role
in describing moments
of inertia of rotational bands in axially deformed nuclei
as those considered in this study.
In addition, as in previous mapped $sdf$-IBM
studies the term proportional to
$(d^\+\times\tilde d)^{(1)}\cdot(f^\+\times\tilde f)^{(1)}$,
arising in the product $\hat L\cdot\hat L$ is omitted
for simplicity.

The energy surface is obtained in terms of the three deformations
$(\beta_1,\beta_2,\beta_3)$ as the expectation value
of the $spdf$-IBM Hamiltonian \eqref{eq:bh}:
\begin{align}
E_{\rm IBM}(\beta_1,\beta_2,\beta_3)
=\frac{\bra{\phi(\beta_1,\beta_2,\beta_3)}\hat H_{\rm B}\ket{\phi(\beta_1,\beta_2,\beta_3)}}
{\braket{\phi(\beta_1,\beta_2,\beta_3)|\phi(\beta_1,\beta_2,\beta_3)}} \; .
\end{align}
Here the ket $\ket{\phi(\beta_1,\beta_2,\beta_3)}$
represents the coherent state \cite{ginocchio1980}
of $s$, $p$, $d$, and $f$ bosons, and is given by
\begin{eqnarray}
 \ket{\phi(\beta_1,\beta_2,\beta_3)}
=\frac{1}{\sqrt{n!}}(\lambda^\+)^{n}\ket{0} \; ,
\end{eqnarray}
where
$\ket{0}$ and $n$ represents the inert core, and
the number of bosons, respectively, and
\begin{eqnarray}
\lambda^\+
=s^\+ + \bar\beta_1 p_0^\+ + \bar\beta_2 d_0^\+ + \bar\beta_3 f_0^\+ \; .
\end{eqnarray}
$\bar\beta_{\lambda}$ stand for the boson analogs of
the dipole, quadrupole, and octupole deformations.
As conventionally made, the bosonic deformations
for each value of $\lambda$ are assumed
to be proportional to the fermionic deformations
\cite{ginocchio1980,nomura2008,nomura2013oct},
i.e.,
\begin{eqnarray}
 \bar\beta_{\lambda} = C_{\lambda} \beta_{\lambda} \; .
\end{eqnarray}
The proportionality constants $C_{\lambda}$ are usually
larger than unity, that is, the bosonic deformation is larger
than the fermionic deformation.
This accounts for the fact that the IBM space is comprised only
of a limited number of valence nucleons, whereas
in the present HFB all constituent nucleons are involved.
The energy surface
$E_{\rm IBM}(\beta_{1},\beta_{2},\beta_{3})$
is obtained as the following analytical expression:
\begin{widetext}
\begin{align}
\label{eq:pes}
E_{\rm IBM}(\beta_{1},\beta_{2},\beta_{3})
&=
\frac{n}{1+\bar\beta_{1}^{2}+\bar\beta_{2}^{2}+\bar\beta_{3}^{2}}
\left(
\epsilon_{s}^{\prime}+
\epsilon_{p}^{\prime}\bar\beta_{1}^{2}+
\epsilon_{d}^{\prime}\bar\beta_{2}^{2}+
\epsilon_{f}^{\prime}\bar\beta_{3}^{2}
\right)
+
\frac{n(n-1)}{(1+\bar\beta_{1}^{2}+\bar\beta_{2}^{2}+\bar\beta_{3}^{2})^2}
\nonumber\\
&
\times
\left[
\kappa_{2}
\left(
2\bar\beta_{2}
-\sqrt{\frac{2}{7}}\chi_{dd}\bar\beta_{2}^{2}
+\sqrt{\frac{2}{3}}\chi_{pp}\bar\beta_{1}^{2}
-2\sqrt{\frac{3}{7}}\chi_{pf}\bar\beta_{1}\bar\beta_{3}
+\frac{2}{\sqrt{21}}\chi_{ff}\bar\beta_{3}^{2}
\right)^{2}
\right.
\nonumber\\
&
\left.
+4\kappa_{3}
\left(
\bar\beta_{3}
+\sqrt{\frac{3}{5}}\chi_{pd}\bar\beta_{1}\bar\beta_{2}
-\frac{2}{\sqrt{15}}\chi_{df}\bar\beta_{2}\bar\beta_{3}
\right)^{2}
+4\kappa_{1}
\left(
\bar\beta_{1}
-\sqrt{\frac{2}{5}}\chi'_{pd}\bar\beta_{1}\bar\beta_{2}
+\frac{3}{\sqrt{35}}\chi'_{df}\bar\beta_{2}\bar\beta_{3}
\right)^{2}
\right] \; .
\end{align} 
\end{widetext}
Note that $\epsilon'$s in the above formula arise from contractions
made when calculating the expectation values of two-body boson terms,
and are defined as
\begin{align}
\label{eq:eps-prime}
&\epsilon_{s}^{\prime}=5\kappa_{2}-7\kappa_3-3\kappa_1
\nonumber\\
&
\epsilon_{d}^{\prime}
=\epsilon_{d}
+6\kappa^{\prime}+(1+\chi_{dd}^2)\kappa_{2}
\nonumber\\
&
\quad
+\frac{7}{5}\kappa_{3}(\chi_{pd}^2+\chi_{df}^2)
+\frac{3}{5}\kappa_{1}(\chi^{\prime\,2}_{pd}+\chi^{\prime\,2}_{df})
\nonumber \\
&
\epsilon_{f}^{\prime}
=-\epsilon_{f} + 12\kappa'
+\frac{5}{7}\kappa_{2}(\chi_{ff}^{2}+\chi_{pf}^2)
\nonumber\\
&
\quad
+(1+\chi_{df}^2)\kappa_{3}
+\frac{3}{7}\kappa_1 {\chi^{\prime\,2}_{df}}
\nonumber \\
&
\epsilon_{p}^{\prime}
=-\epsilon_{p}
+\frac{5}{3}\kappa_{2}(\chi_{pp}^{2}+\chi_{pf}^2)
+(1+{\chi^{\prime\,2}_{pd}})\kappa_{1}
+\frac{7}{3}\kappa_3 \chi_{df}^2 \; .
\end{align}

The procedure to determine
the parameters that appear in \eqref{eq:pes} including
the coefficients $C_{\lambda}$ is summarized as follows.
\begin{enumerate}[label=(\roman{enumi})]
 \item
The HFB-to-IBM mapping is carried out
in the $(\beta_2,\beta_3)$-deformation space with $\beta_1=0$.
First, the parameters ($\epsilon_d$, $\kappa_2$, $\chi_{dd}$, $C_2$),
which are related to $sd$ bosons,
are obtained by reproducing the topology near the minimum
of the HFB PES along the $\beta_3=0$ axis, that is,
the deformation at which
the minimum occurs, depth and steepness of the potential valley,
should be reasonably reproduced.
Since the $\hat L\cdot\hat L$ term does not make a
unique contribution to the energy surface, the strength
$\kappa'$ has to be determined separately in such a way
\cite{nomura2011rot}
that the cranking moment of inertia calculated in
the intrinsic state of the $sd$-boson system
at the minimum on the $\beta_3=0$ axis
should be equal to the Thouless-Valatin moment
of inertia \cite{TV} calculated by the HFB.

\item
Second, the parameters that are related to $f$ bosons
(or octupole deformation),
i.e., ($\epsilon_f$, $\chi_{ff}$, $\kappa_3$,
$\chi_{df}$, $C_3$)
are determined so that the characteristic features near
the minimum of the HFB $(\beta_2,\beta_3)$-PES,
that is, the location of the minimum,
and steepness in both the $\beta_2$ and $\beta_3$
deformations, are reproduced as closely as possible.
A more detailed account of the steps (i) and (ii)
of the mapping procedure is found
in Ref.~\cite{nomura2020oct}.
Constraining to the zero dipole moment $\beta_1=0$
in these steps is due to the fact that
the energy surfaces are significantly soft
against the $\beta_1$ deformation (see Sec.~\ref{sec:pes}),
and is in order to exploit the $sdf$-IBM
parameters obtained from the previous
study of Ref.~\cite{nomura2020oct}.

\item
In the final step, the parameters
($\epsilon_p$, $\chi_{pf}$, $\chi_{pp}$, $\chi_{pd}$,
$\kappa_1$, $\chi_{df}'$, $\chi_{pd}'$, $C_1$)
are determined so that the bosonic PESs in the
$(\beta_1,\beta_2)$ and $(\beta_1,\beta_3)$,
with the $\beta_3$ and $\beta_2$ deformations being
fixed at those values corresponding to the
minimum in the $(\beta_2,\beta_3)$-deformation
space, respectively, should be as similar as
possible to the corresponding HFB PESs.
Here it is assumed for simplicity that
$\epsilon_p'=\epsilon_f'$ in Eq.~\eqref{eq:eps-prime},
and that the strength of the dipole-dipole interaction
equals that of the octupole-octupole one, i.e.,
$\kappa_1=\kappa_3$.
In addition, the fixed values of the parameters $\chi_{pf}=-0.5$,
$\chi_{pd}=0.2$, and $\chi_{pd}'=0.5$
are used to reflect the fact that, as shown below,
the PESs are not very sensitive to the $\beta_1$
deformation. Also their values are chosen to be
smaller in magnitude than $\chi_{ff}$, $\chi_{df}$,
and $\chi_{df}'$ parameters,
assuming that couplings of $p$ bosons to
$d$ and $f$ bosons are weak.
The remaining parameters ($\chi_{pp}$, $\chi_{df}'$, $C_1$)
are chosen to vary gradually as functions of boson number.
\end{enumerate}

The mapped $spdf$-IBM Hamiltonian is numerically
diagonalized \cite{arbmodel} in the space consisting of
$n(=n_s+n_p+n_d+n_f)$ bosons,
where $n_s$, $n_p$, $n_d$, and $n_f$ are numbers of
$s$, $p$, $d$, and $f$ bosons, respectively.
As in the standard IBM, the total number of bosons
$n$ is conserved for each nucleus.
Also the negative-parity $p$ and $f$ bosons are treated
on the same footing as the positive-parity $s$ and $d$ bosons,
that is, the numbers $n_s$, $n_p$, $n_d$, and $n_f$
are all allowed to take values from 0 to $n$.

The $E2$, $E3$, and $E1$
transition probabilities are computed
by using the corresponding transition operators
\begin{eqnarray}
\label{eq:tel}
\hat T^{(E2)} = e_{2} \hat Q
\; , \quad
\hat T^{(E3)} = e_{3} \hat O
\; , \quad
\hat T^{(E1)} = e_{1} \hat D' \; ,
\end{eqnarray}
where $\hat Q$ and $\hat O$ are the same quadrupole
and octupole boson operators in Eq.~\eqref{eq:q}
and Eq.~\eqref{eq:o}, respectively, with the
same values of the parameters $\chi$ as those
used for the Hamiltonian.
It was shown in Ref.~\cite{otsuka1988} that the dipole transition
operator should be expressed with parameters
that exhibit some dependence on boson number,
which is on the basis of the fact that
the $E1$ transitions are accounted for
to a large extent by the single-particle degrees of
freedom or underlying shell structure.
The $E1$ operator introduced in \cite{otsuka1988} is adopted
here for $\hat D'$ in \eqref{eq:tel} in the $spdf$-boson model,
and is given as
\begin{align}
\label{eq:e1}
 \hat D'=
&\,s^\+\tilde p + p^\+\tilde s
-\{1-\alpha(n-7)\}
\nonumber\\
&\times
\left[
(p^\+\times\tilde d + d^\+\times\tilde p)^{(1)}
-\beta (f^\+\times\tilde d + d^\+\times\tilde f)^{(1)}
\right] \; ,
\end{align}
with the parameters $\alpha=0.4$ and $\beta=0.5$ \cite{otsuka1988}.
For the $sdf$-boson model, $\hat D'$ is simply
given as
\begin{align}
\label{eq:e1-sdf}
\hat D'=(d^\+\times\tilde f + f^\+\times\tilde d)^{(1)} \; . 
\end{align}
The boson effective charges $e_2$ and $e_3$
for the $E2$ and $E3$ operators are determined \cite{nomura2020oct}
for each nucleus so that the expectation values
$\braket{\hat Q}$ and $\braket{\hat O}$ in the coherent state
should be equal to the $q_2^{\rm min}$ and $q_3^{\rm min}$ moments
obtained in the HFB that correspond to the minimum
in the $(\beta_2,\beta_3)$-PES, that is,
\begin{eqnarray}
\label{eq:ech-q}
&e_{2} = e_{2}^{(0)}q_{2}^{\rm min} (\braket{\hat Q}_{\rm min})^{-1} \\
\label{eq:ech-o}
&e_{3} = e_{3}^{(0)}q_{3}^{\rm min} (\braket{\hat O}_{\rm min})^{-1} \; .
\end{eqnarray}
Note that the multiplication by additional
overall factors $e_{2}^{(0)}=0.12$ $e$b
and $e_{3}^{(0)}=0.06$ $e$b$^{3/2}$ in the above formulas
is in order that the
calculated $B(E2;2^+_1 \to 0^+_1)$ and $B(E3;3^-_1 \to 0^+_1)$
should be of the same order of magnitudes as
the experimental values.
In addition, common $E2$ and $E3$ effective charges, which are
determined by \eqref{eq:ech-q} and \eqref{eq:ech-o},
are used for the $sdf$- and $spdf$-IBM.
The $E1$ boson charge $e_1$ is fixed to be the value
employed in Ref.~\cite{otsuka1988}, i.e., $e_1=0.0054$ $e$b$^{1/2}$,
in the $spdf$-boson model, while the constant value
$e_1=0.008$ $e$b is adopted in the $sdf$-boson model.
Also for the $spdf$-model, in order to see if the parametrization
of the operator in \eqref{eq:e1} is adequate
the $E1$ transitions are computed
by using $\hat D'$
that is equal to $\hat D$ \eqref{eq:d} in the Hamiltonian
with the constant effective charge $e_1=0.008$ $e$b$^{1/2}$
as in the $sdf$-IBM.

%
%
\begin{figure*}[ht]
\begin{center}
\includegraphics[width=.49\linewidth]{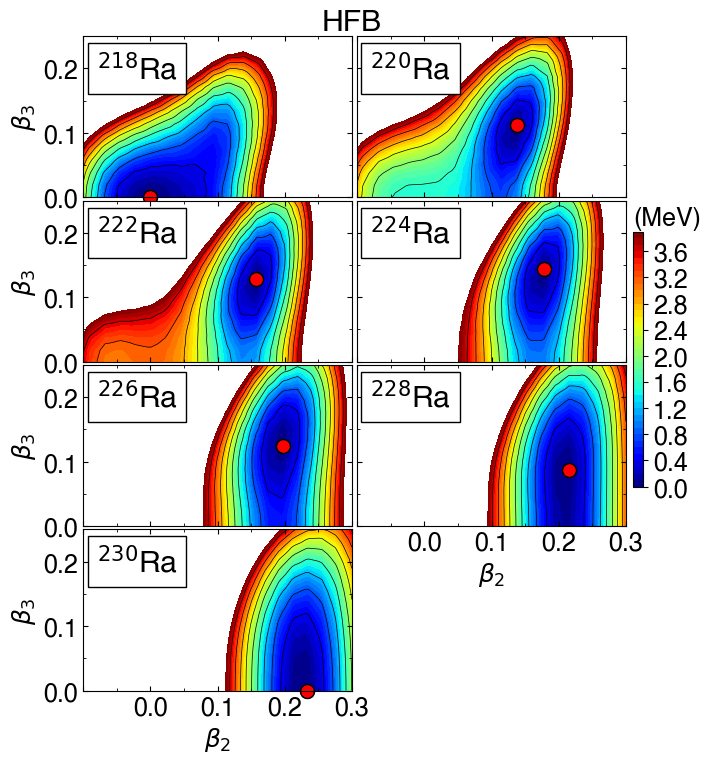}
\includegraphics[width=.49\linewidth]{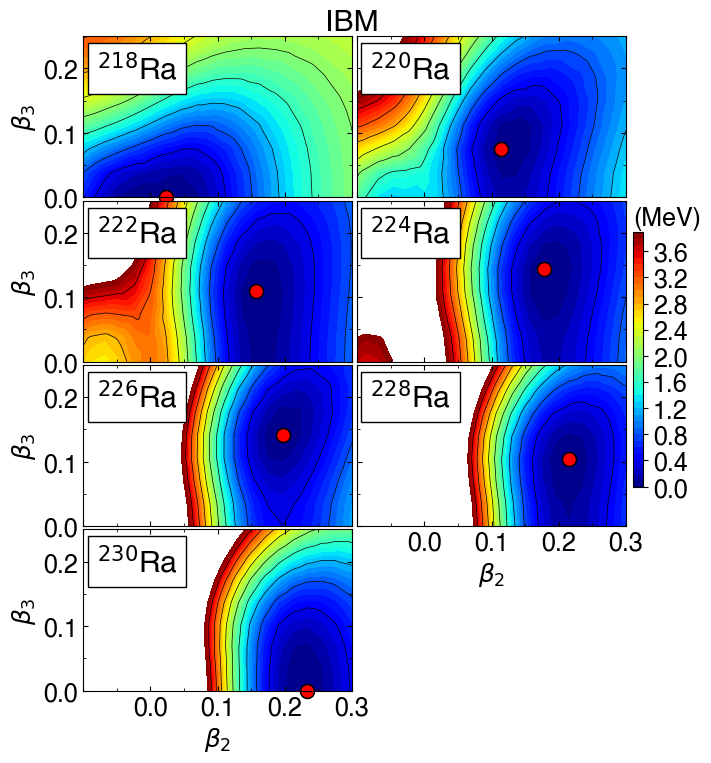}
\caption{
Potential energy surfaces for $^{218-230}$Ra in terms of
the axial quadrupole $\beta_2$ and octupole $\beta_3$ deformations
obtained from the HFB method using the
Gogny-D1M EDF with the dipole deformation set to $\beta_1=0$,
and the corresponding energy surfaces in the $spdf$-IBM.
The minimum is indicated by the solid circle,
and energy difference between neighboring contours
is 0.4 MeV.}
\label{fig:b23}
\end{center}
\end{figure*}

%
%
\begin{figure*}[ht!]
\begin{center}
\includegraphics[width=.49\linewidth]{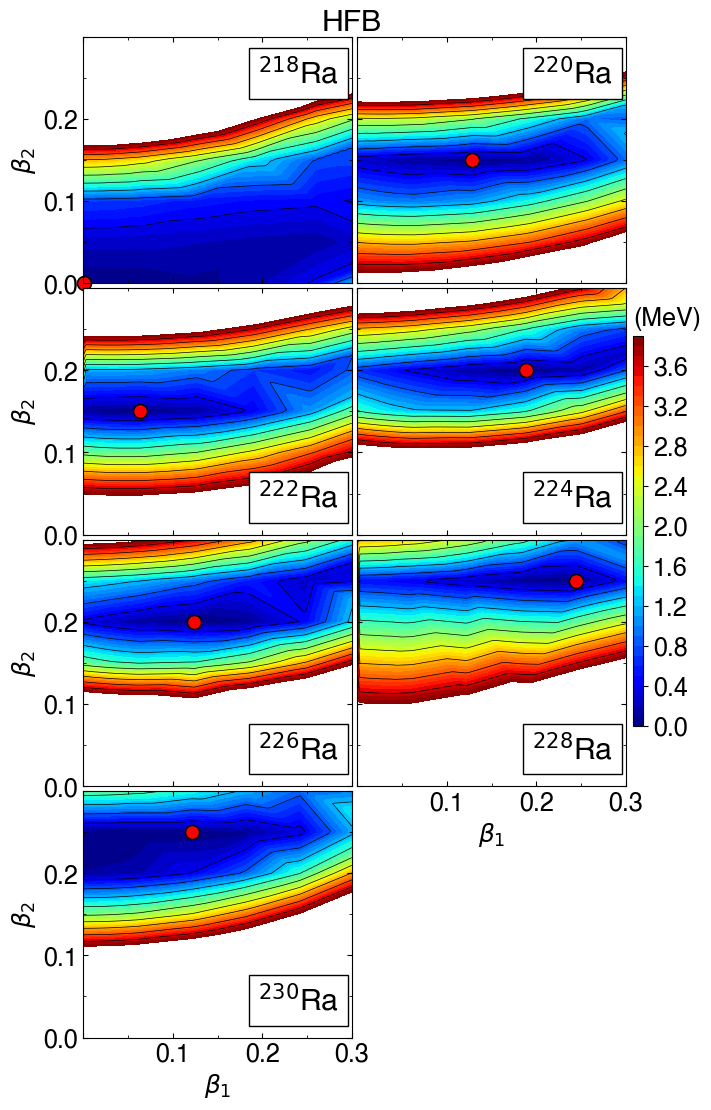}
\includegraphics[width=.49\linewidth]{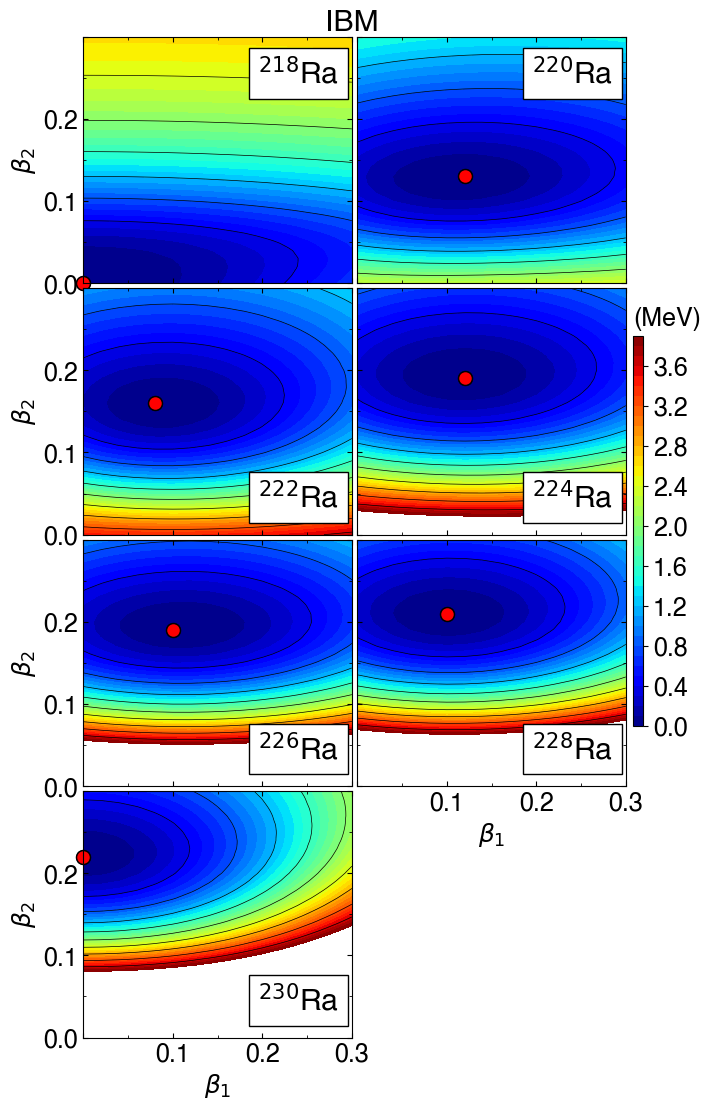}
\caption{Same as the caption to Fig.~\ref{fig:b23},
but for the dipole $\beta_1$ and quadrupole $\beta_2$
deformations with the octupole deformation $\beta_3$
fixed to be the value corresponding to
the minimum in the $(\beta_2,\beta_3)$ space with $\beta_1=0$.}
\label{fig:b12}
\end{center}
\end{figure*}

%
%
\begin{figure*}[ht!]
\begin{center}
\includegraphics[width=.49\linewidth]{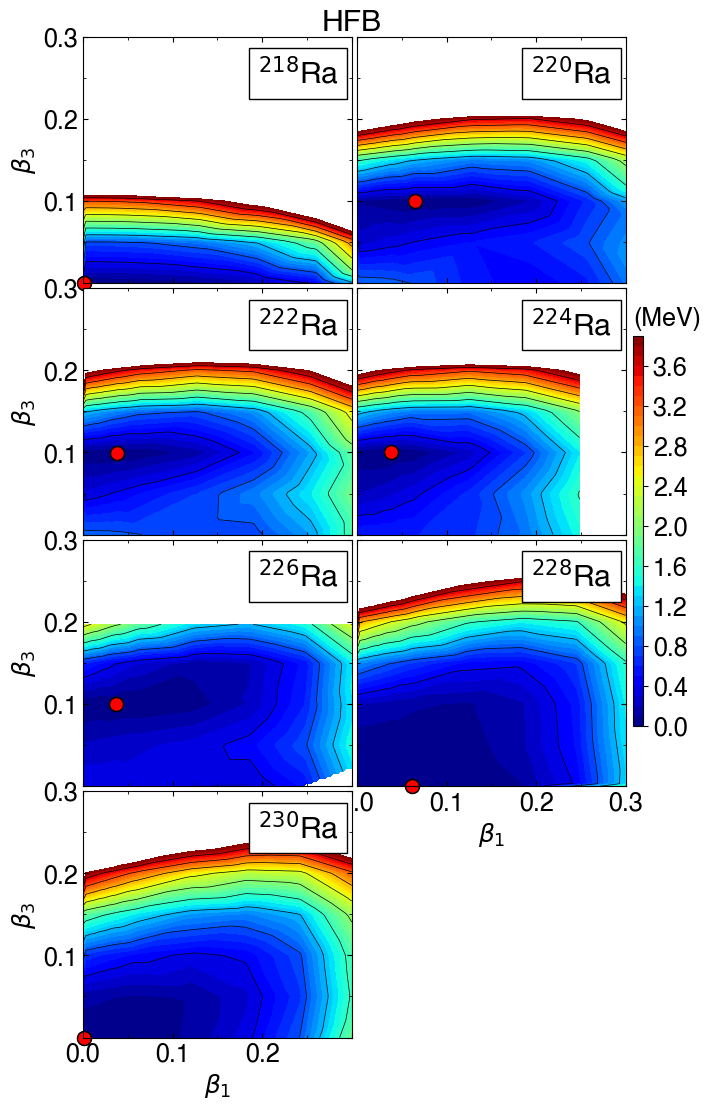}
\includegraphics[width=.49\linewidth]{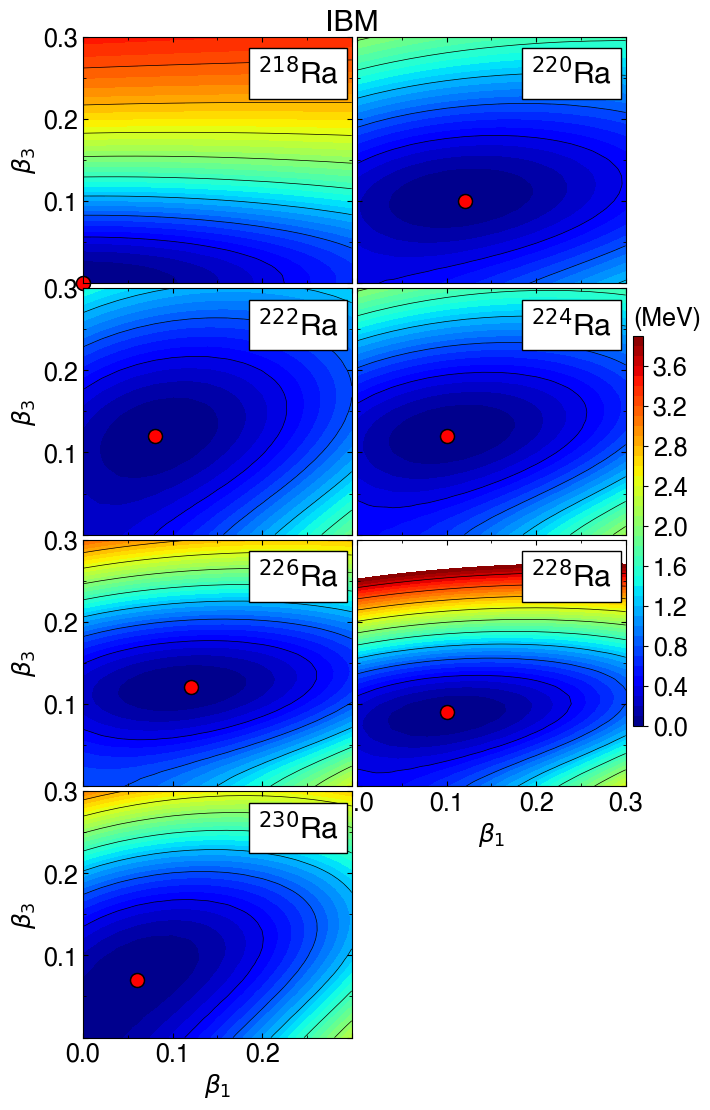}
\caption{Same as the caption to Fig.~\ref{fig:b23},
but for the dipole $\beta_1$ and octupole $\beta_3$
deformations with the quadrupole deformation $\beta_2$
fixed to be the value corresponding to
the minimum in the $(\beta_2,\beta_3)$ space with $\beta_1=0$.}
\label{fig:b13}
\end{center}
\end{figure*}

%
%
\begin{figure}[ht!]
\begin{center}
\includegraphics[width=\linewidth]{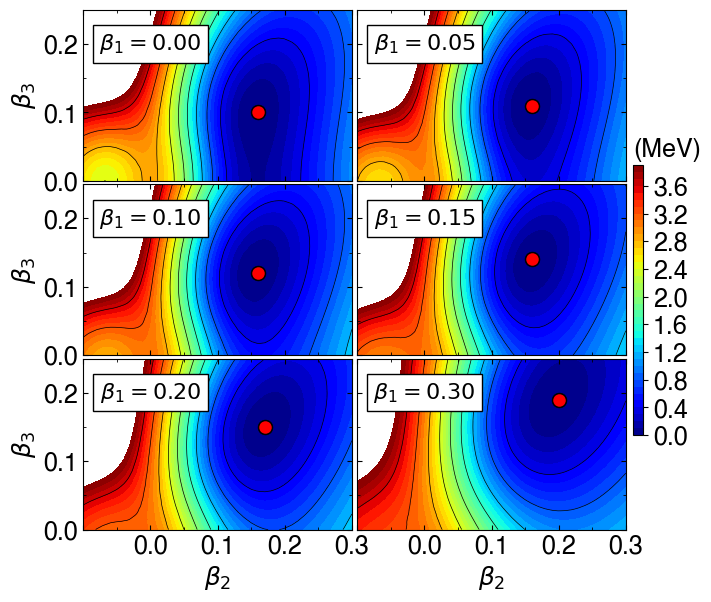}
\caption{Mapped $spdf$-IBM $(\beta_2,\beta_3)$-PESs for
$^{222}$Ra with the $\beta_1$ deformation being
fixed at several different values.}
\label{fig:b23-p}
\end{center}
\end{figure}

\section{Results\label{sec:results}}

\subsection{Potential energy surfaces\label{sec:pes}}

Figures~\ref{fig:b23}, \ref{fig:b12}, and \ref{fig:b13}
show the two-dimensional PESs in the intrinsic $(\beta_2,\beta_3)$,
$(\beta_1,\beta_2)$, and $(\beta_1,\beta_3)$ deformations,
respectively, computed for the $^{218-230}$Ra isotopes
using the constrained HFB method,
and the corresponding mapped IBM PESs.
The results for the Th isotopes are very similar
to those for the Ra isotopes, and are therefore not shown.
Note also that the $(\beta_2,\beta_3)$-PESs in Fig.~\ref{fig:b23}
are also shown in Fig.~1 of Ref.~\cite{nomura2020oct}.

The HFB $(\beta_2,\beta_3)$-PESs exhibit a pronounced octupole
minimum for those Ra and Th nuclei with $N=134$ or 136.
For $^{226}$Ra and $^{228}$Ra, shown in Fig.~\ref{fig:b23},
the potential is steep
along the $\beta_2$ deformation, and is substantially soft
in $\beta_3$ deformation.
The behavior of the HFB PES for the Ra and Th nuclei
with increasing neutron number points to a shape phase
transition from nearly spherical to stable octupole
deformed, and to octupole-soft intrinsic shapes
\cite{nomura2013oct,nomura2014,nomura2020oct}.
The mapped IBM PESs follow the nucleon-number dependence
and overall topology of the HFB PESs.
The bosonic PESs are generally softer than the fermionic ones
for those configurations corresponding to large deformations
that are far from the minimum.
This is a common feature of the mapping procedure,
and is explained by the fact that the boson model space
is rather limited in comparison with the SCMF model.
In the mean-field configurations at large deformations
single(quasi)particle excitations become important,
which are not included in the usual IBM space.
The mapping is considered valid, as long as it is
carried out specifically in the vicinity of the global minimum
of the HFB PES, since the configurations near the
global minimum are most relevant to the low-energy
collective states.
More detailed accounts of the mapping procedure
are found in
Refs.~\cite{nomura2008,nomura2013oct,nomura2014,nomura2020oct}.

The $(\beta_1,\beta_2)$-PESs obtained from the HFB calculation
are in general steep in $\beta_2$ deformation, but
are considerably soft along the $\beta_1$ deformation.
Their global features also do not depend
significantly on nucleon numbers, except for that
of the nearly spherical nucleus $^{218}$Ra.
These observations indicate that the
intrinsic shape of the Ra and Th nuclei is not
sensitive to the dipole deformation.
The HFB calculation also suggests
that, for many of the Ra nuclei, a very shallow minimum occurs
at finite $\beta_1$ deformation, as seen in Fig.~\ref{fig:b12}.
The corresponding IBM PESs exhibit similar systematic
behaviors as the HFB PESs.
The IBM PESs are constructed so as to reproduce
mainly the location of the $\beta_1\neq0$ minimum,
and the softness in the $\beta_1$ deformation.

The HFB PESs in the $(\beta_1,\beta_3)$ plane, shown
in Fig.~\ref{fig:b13}, are $\beta_1$ soft, and more or
less resemble a flat-bottomed potential
in $\beta_3$ deformation particularly for $^{228}$Ra and $^{230}$Ra.
A shallow $\beta_1$ minimum is observed in the HFB PESs
for $^{220}$Ra, $^{222}$Ra, $^{224}$Ra, and $^{226}$Ra.
The IBM mapping is carried out so that the
softness in $\beta_1$ and $\beta_3$ deformations,
as well as the location of the minimum, of the HFB PES
is reproduced as much as possible.

Some deviations in topology from the original
HFB PESs are observed in the
$(\beta_1,\beta_2)$ and $(\beta_1,\beta_3)$ planes.
For instance, for $^{228}$Ra and $^{230}$Ra the mapped IBM
$(\beta_1,\beta_3)$-PESs exhibits a $\beta_3\neq0$
minimum, while in the HFB PESs the minimum occurs
on the $\beta_3=0$ axis.
The deviations can be accounted for by the limited analytical
form of the $spdf$-IBM, which does not have
degrees of freedom sufficient to
reproduce accurately the HFB PESs.
In addition, the $spdf$-IBM parameters relevant
to the dipole deformations are chosen to reproduce
the global features of both the
HFB $(\beta_1,\beta_2)$- and $(\beta_1,\beta_3)$-PESs,
and it is not very obvious to
reproduce all the details of the two surfaces
simultaneously.

To see possible effects of
the $p$ boson degree of freedom on
the energy surface,
Fig.~\ref{fig:b23-p} depicts the
$spdf$-IBM $(\beta_2,\beta_3)$-PESs for $^{222}$Ra,
with the $\beta_1$ deformation being
fixed at several different values.
$^{222}$Ra is taken here as an illustrative case,
since as shown below the inclusion of $p$
bosons has pronounced influences on
energy spectra in this nucleus.
One can see from Fig.~\ref{fig:b23-p}
that, with the increasing $\beta_1$ deformation,
the $(\beta_2,\beta_3)$-PES becomes steeper
in $\beta_3$, and the minimum occurs at
larger $\beta_2$ and $\beta_3$ deformations.
For a relatively small dipole
deformation of $\beta_1=0.05$,
near which the equilibrium minimum is
found in the $(\beta_1,\beta_2)$- and
$(\beta_1,\beta_3)$-PESs, $p$ bosons do
not appear to have significant effects
of altering the topology of the
$(\beta_2,\beta_3)$-PES.

%
%
\begin{table*}[hbt]
\caption{\label{tab:para}
Parameters for the $spdf$-IBM Hamiltonian adopted in
the present work for the nuclei $^{218-230}$Ra and $^{220-232}$Th.
$\epsilon_d$, $\epsilon_f$, $\epsilon_p$,
$\kappa_2$, $\kappa'$, $\kappa_3$ are in keV units.
Fixed values $\chi_{pf}=-0.5$,
$\chi_{pd}=0.2$, and $\chi'_{pd}=0.5$
are used, and the dipole-dipole
interaction strength is assumed to be equal to
the octupole-octupole one, $\kappa_1=\kappa_3$.
}
\begin{center}
 \begin{ruledtabular}
\begin{tabular}{ccccccccccccccc}
\textrm{} &
\textrm{$\epsilon_d$} &
\textrm{$\epsilon_f$}&
\textrm{$\epsilon_p$}&
\textrm{$\kappa_2$} &
\textrm{$\chi_{dd}$}&
\textrm{$\chi_{pp}$}&
\textrm{$\chi_{ff}$} &
\textrm{$\kappa'$}&
\textrm{$\kappa_3$}&
\textrm{$\chi_{df}$}&
\textrm{$\chi'_{df}$}&
\textrm{$C_2$}&
\textrm{$C_3$}&
\textrm{$C_1$}
\\
\hline
$^{218}$Ra & $363$ & $-672$ & $-701$ & $-40$ & $-0.50$ & $1.50$ & $1.20$ & $0$ & $-26$ & $-1.20$ & $0.60$ & $10.0$ & $7.0$ & $1.8$ \\ 
$^{220}$Ra & $533$ & $-789$ & $-505$ & $-53$ & $-1.30$ & $1.50$ & $1.70$ & $-15$ & $-26$ & $-2.20$ & $1.10$ & $7.6$ & $7.0$ & $3.2$ \\ 
$^{222}$Ra & $213$ & $-526$ & $-241$ & $-52$ & $-1.30$ & $0.90$ & $1.90$ & $-11$ & $-22$ & $-1.90$ & $0.95$ & $7.2$ & $5.6$ & $3.1$ \\ 
$^{224}$Ra & $266$ & $-476$ & $-130$ & $-52$ & $-1.30$ & $0.50$ & $1.80$ & $-14$ & $-20$ & $-2.10$ & $1.10$ & $6.5$ & $6.8$ & $2.5$ \\ 
$^{226}$Ra & $318$ & $-755$ & $-414$ & $-46$ & $-1.30$ & $0.50$ & $1.80$ & $-11$ & $-19$ & $-2.60$ & $1.10$ & $6.3$ & $6.9$ & $2.0$ \\ 
$^{228}$Ra & $264$ & $-800$ & $-512$ & $-44$ & $-1.30$ & $0.50$ & $1.80$ & $-10$ & $-19$ & $-2.20$ & $1.10$ & $5.8$ & $8.0$ & $2.0$ \\ 
$^{230}$Ra & $207$ & $-738$ & $-530$ & $-41$ & $-1.30$ & $0.50$ & $1.60$ & $-8$ & $-18$ & $-1.80$ & $0.90$ & $5.5$ & $4.9$ & $2.0$ \\ 
$^{220}$Th & $619$ & $-899$ & $-940$ & $-49$ & $-0.30$ & $1.50$ & $1.20$ & $0$ & $-27$ & $-1.20$ & $0.60$ & $7.3$ & $4.0$ & $2.0$ \\ 
$^{222}$Th & $475$ & $-960$ & $-657$ & $-53$ & $-1.10$ & $1.50$ & $1.90$ & $-17$ & $-27$ & $-1.90$ & $0.95$ & $6.7$ & $6.2$ & $2.8$ \\ 
$^{224}$Th & $276$ & $-398$ & $-151$ & $-50$ & $-1.34$ & $0.90$ & $1.68$ & $-11$ & $-24$ & $-1.75$ & $0.88$ & $6.8$ & $7.0$ & $2.7$ \\ 
$^{226}$Th & $356$ & $-600$ & $-254$ & $-53$ & $-1.32$ & $0.70$ & $1.80$ & $-15$ & $-22$ & $-2.00$ & $1.00$ & $6.1$ & $5.5$ & $2.2$ \\ 
$^{228}$Th & $482$ & $-739$ & $-364$ & $-51$ & $-1.32$ & $0.50$ & $1.82$ & $-16$ & $-16$ & $-2.40$ & $1.20$ & $5.6$ & $5.0$ & $2.0$ \\ 
$^{230}$Th & $418$ & $-683$ & $-362$ & $-49$ & $-1.32$ & $0.50$ & $1.80$ & $-15$ & $-16$ & $-1.90$ & $0.95$ & $5.3$ & $5.1$ & $1.8$ \\ 
$^{232}$Th & $373$ & $-657$ & $-400$ & $-47$ & $-1.30$ & $0.50$ & $1.70$ & $-13$ & $-16$ & $-1.50$ & $0.75$ & $5.1$ & $4.4$ & $2.0$ \\
\end{tabular}
 \end{ruledtabular}
\end{center}
\end{table*}

\subsection{IBM parameters\label{sec:para}}

The $spdf$-IBM parameters employed in the present work
are given in Table~\ref{tab:para}.
The parameters
($\epsilon_d$, $\kappa_2$, $\chi_{dd}$, $C_2$,
$\epsilon_f$, $\chi_{ff}$, $\kappa_3$, $\chi_{df}$, $C_3$)
for the $sdf$-IBM Hamiltonian are also found
in Fig.~3 of Ref.~\cite{nomura2020oct}.
The $d$-boson energy basically decreases with $N$,
as the quadrupole collectivity enhances.
As in the previous mapped $sdf$-IBM calculations
the single-$f$ boson energy $\epsilon_f$ exhibits
a parabolic behavior, reaching a minimum in magnitude
near $N=136$ (Ra) and $N=134$ (Th), in which
the $(\beta_2,\beta_3)$-PES exhibits a pronounced
octupole deformation.
What is of particular interest is the fact that
the $p$-boson energy $\epsilon_p$ is here systematically
lower in magnitude than the $f$-boson one, and is
of the same order of magnitude as or
even lower than the $d$-boson energy for many of the nuclei
considered.
This is because
the $(\beta_1,\beta_2)$ and $(\beta_1,\beta_3)$ energy
surfaces are in many cases soft in $\beta_1$
deformation, and the mapping procedure incorporates
such features in the parameters (see Figs.~\ref{fig:b12} and \ref{fig:b13}).
The $\kappa_2$ and $\kappa_3(=\kappa_1)$ parameters
are almost constant but shows
a gradual decreases in magnitude with neutron number.
The derived parameter $\chi_{dd}$ is close to
the SU(3) value $-\sqrt{7}/2$ for most of the nuclei
with $N\geqslant134$, which exhibit a strong prolate deformation.

The $\chi_{pp}$ parameter exhibits
a substantial variation
from $N=132-136$.
For $N=130$ and 132, the chosen $\chi_{pp}$ values
are of the same order of magnitude as $\chi_{ff}$,
indicating that the dipole correlations can be as pronounced
as the octupole correlations in these transitional nuclei.
For the nuclei with $N=130$ and 132,
the large $\chi_{pp}$ values are required
to produce the $\beta_1$ softness of the
$(\beta_1,\beta_2)$- and $(\beta_1,\beta_3)$-PESs
with limited boson numbers.
Both $\chi_{ff}$ and $\chi_{df}$ parameters concern
the degree of octupole deformation, and becomes maximal
for those nuclei near $N=134$ or 136,
for which the octupole minimum is most pronounced.
The values of the $\chi_{df}'$ parameter are
chosen to be $\chi'_{df}\approx\chi_{df}/2$, for simplicity.
Also the absolute values $|\chi_{df}|$ and $|\chi_{df}'|$
are generally larger than $|\chi_{pd}|=0.2$ and $|\chi_{pd}'|=0.5$,
respectively,
assuming that the coupling of $p$ to $d$ bosons
is weaker than that of $f$ to $d$ bosons.

The scale factor $C_2$ for the $\beta_2$ deformation
only exhibits a gradual decrease with neutron number.
The factor $C_3$ used in Ref.~\cite{nomura2020oct}
is here slightly modified so that the minima
in all the $(\beta_2,\beta_3)$,
$(\beta_1,\beta_2)$ and $(\beta_1,\beta_3)$ surfaces
are consistently reproduced to a certain accuracy.
This is the reason why this coefficient exhibits
a rather irregular behavior with neutron number.
The $C_1$ coefficient here displays a characteristic
feature of being maximal near $N=132$,
in which the energy surface is particularly
soft in the $(\beta_1,\beta_2)$ and $(\beta_1,\beta_3)$
deformation spaces.

%
%
\begin{figure*}[htb!]
\begin{center}
\includegraphics[width=.7\linewidth]{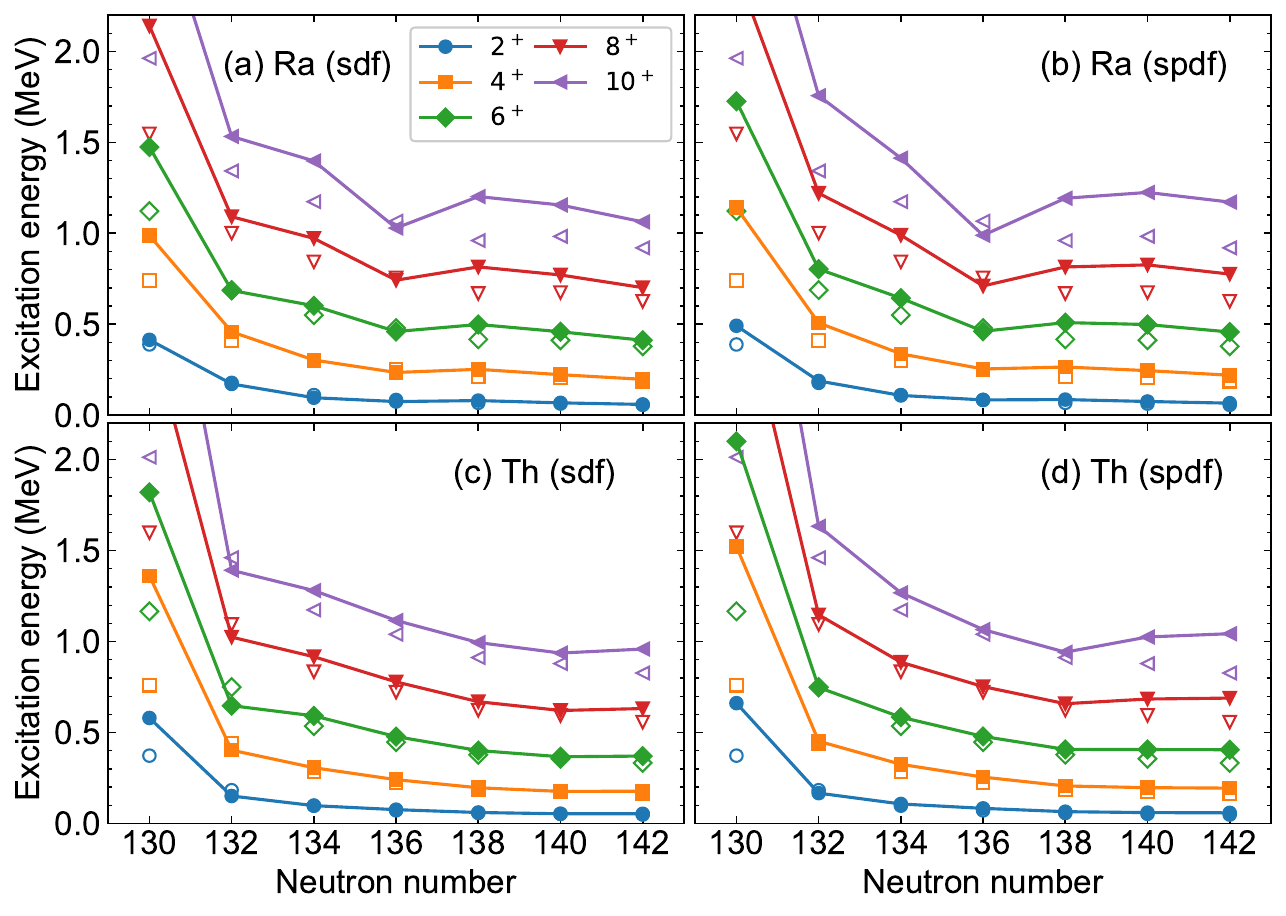}\\
\caption{Calculated excitation energies for even-spin positive-parity
yrast states for $^{218-230}$Ra and $^{220-232}$Th
within the mapped $sdf$-IBM and $spdf$-IBM (solid symbols
connected by lines).
The corresponding experimental values,
represented by open symbols,
are taken from NNDC \cite{data}.}
\label{fig:level-pos}
\end{center}
\end{figure*}

%
%
\begin{figure*}[htb!]
\begin{center}
\includegraphics[width=.7\linewidth]{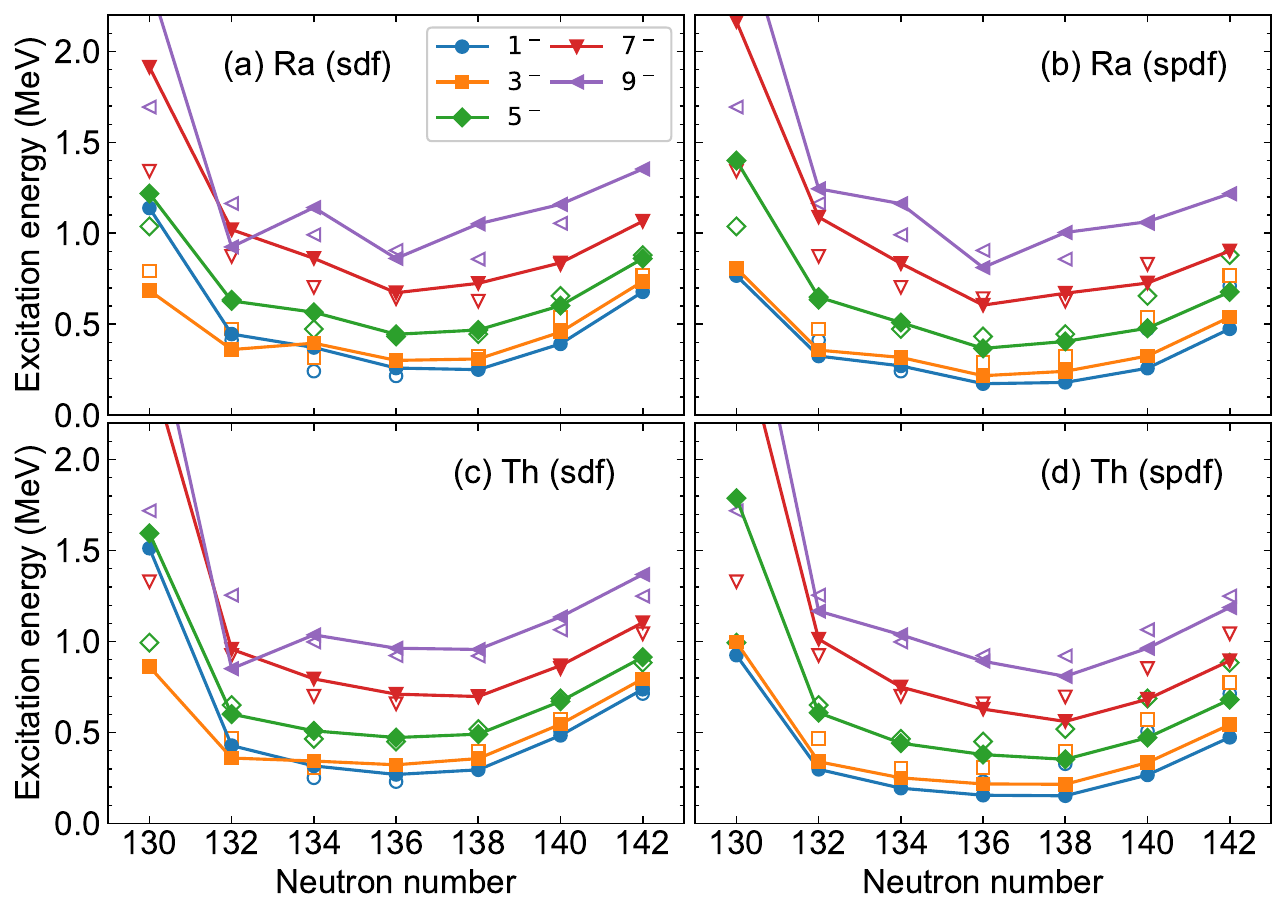}\\
\caption{Same as the caption to Fig.~\ref{fig:level-pos},
but for the odd-spin negative-parity yrast states.}
\label{fig:level-neg}
\end{center}
\end{figure*}

%
%
\begin{figure}[htb!]
\begin{center}
\includegraphics[width=\linewidth]{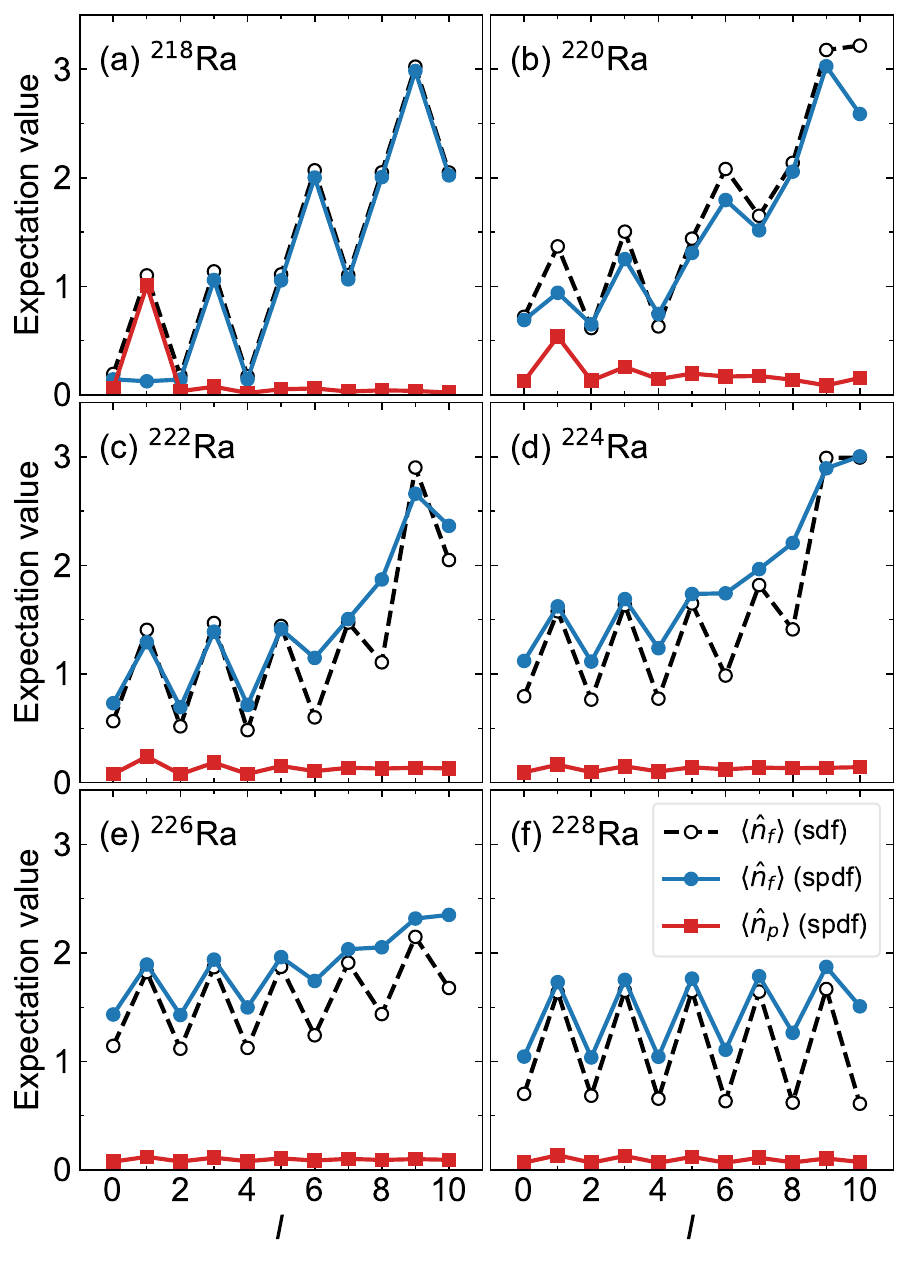}\\
\caption{
Expectation values of the $f$-boson
number operator $\braket{\hat n_f}$ in the $sdf$-IBM
and $spdf$-IBM, and of the $p$-boson
number operator $\braket{\hat n_p}$ in the $spdf$-IBM
for the even-spin positive-parity and odd-spin negative-parity
yrast states with increasing spin $I$
in $^{218-228}$Ra.}
\label{fig:wf}
\end{center}
\end{figure}

\subsection{Low-energy spectra\label{sec:energies}}

Figures~\ref{fig:level-pos} and \ref{fig:level-neg}
display, respectively, the predicted energy spectra
for the positive-parity
yrast states in the ground-state $K=0^+$ band,
and for the negative-parity yrast states in the $K=0^-$
band in the mapped $sdf$-IBM and $spdf$-IBM in comparison
with the experimental data \cite{data}.

The $sdf$- and $spdf$-boson model calculations
both suggest that the positive-parity energy levels gradually
decrease as functions of the neutron number, reflecting
the enhancement of the quadrupole collectivity or
shape phase transitions from the nearly spherical
to strongly prolate deformed regimes.
The two boson models appear to give rise to
qualitatively very similar results.
However, the $spdf$-boson model gives the positive-parity levels
that are rather higher in energy than those in the $sdf$-IBM.
These differences are inferred from the structure
of the calculated wave functions of the relevant states,
as shown in Fig.~\ref{fig:wf}.
For the nuclei $^{222-228}$Ra [Figs.~\ref{fig:wf}(c)-\ref{fig:wf}(d)],
the expectation values of the $f$-boson number
operator $\braket{\hat n_{f}}$ in the wave functions
for the even-spin states in the
ground-state $K=0^+$ band in the $spdf$-IBM
are systematically larger than those in the $sdf$-IBM.
These effects, together with the presence of $p$ bosons,
could account for the quantitative differences
between the $sdf$-IBM and $spdf$-IBM results for
the positive-parity states.

As shown in Fig.~\ref{fig:level-neg},
both the $sdf$-IBM and $spdf$-IBM 
reproduce the observed parabolic behaviors
of the negative-parity energy spectra in $^{218-230}$Ra
and $^{220-232}$Th as functions of the neutron
number, with the bandhead $1^-$ state
of the possible $K=0^-$ band being lowest in energy near $N=136$.
The effects of including $p$ bosons
are such that the $1^-$ bandhead of the $K=0^-$ bands
are systematically lowered.
For those Ra and Th nuclei with $N\leqslant132$,
in the $sdf$-IBM the lowest-energy negative-parity state
is the $3^-$ state, and the $1^-$ level is above that
of the $3^-$ one.
Also the $9^-$ level appears below the $7^-$ level
for $^{220}$Ra and $^{222}$Th in the $sdf$-IBM.
In the $spdf$-IBM results, however, the $1^-$ state is
predicted to be the lowest-lying negative-parity state
for these nuclei, which is consistent with
the experimental data for $^{220}$Ra in particular.
The irregularity in the order of the $7^-$ and $9^-$
at $N=132$ appears to be removed in the $spdf$-IBM.
Overall the $spdf$-IBM underestimates the observed $1^-$ level.
This is, to a large extent, due to the fact that
in the present IBM mapping calculation
the single $p$-boson energy is substantially low
with respect to those of $d$ and $f$ bosons
(cf. Table~\ref{tab:para}).

It is seen from Fig.~\ref{fig:wf} that the contributions
of $f$ bosons to the wave functions for the
odd-spin negative-parity states in the $spdf$-IBM are
within the range $\braket{\hat n_f}\approx1.5-3.0$ particularly
for $^{222-228}$Ra,
which are of the same order of magnitude as those
in the $sdf$-IBM.
The contributions of $p$ bosons are mostly minor
for these nuclei, and are rather constant with the
increasing spin.
These results suggest that the odd-spin negative-parity yrast states
in deformed Th and Ra nuclei are mostly made of $f$ bosons
even in the presence of $p$ bosons.
Only for the $^{218}$Ra nucleus (also for $^{220}$Th),
the $1^-$ wave function
is characterized by the one-$p$-boson configurations,
$\braket{\hat n_p}\approx1$, while the $f$-boson content
is minor.

%
%
\begin{figure}[htb!]
\begin{center}
\includegraphics[width=\linewidth]{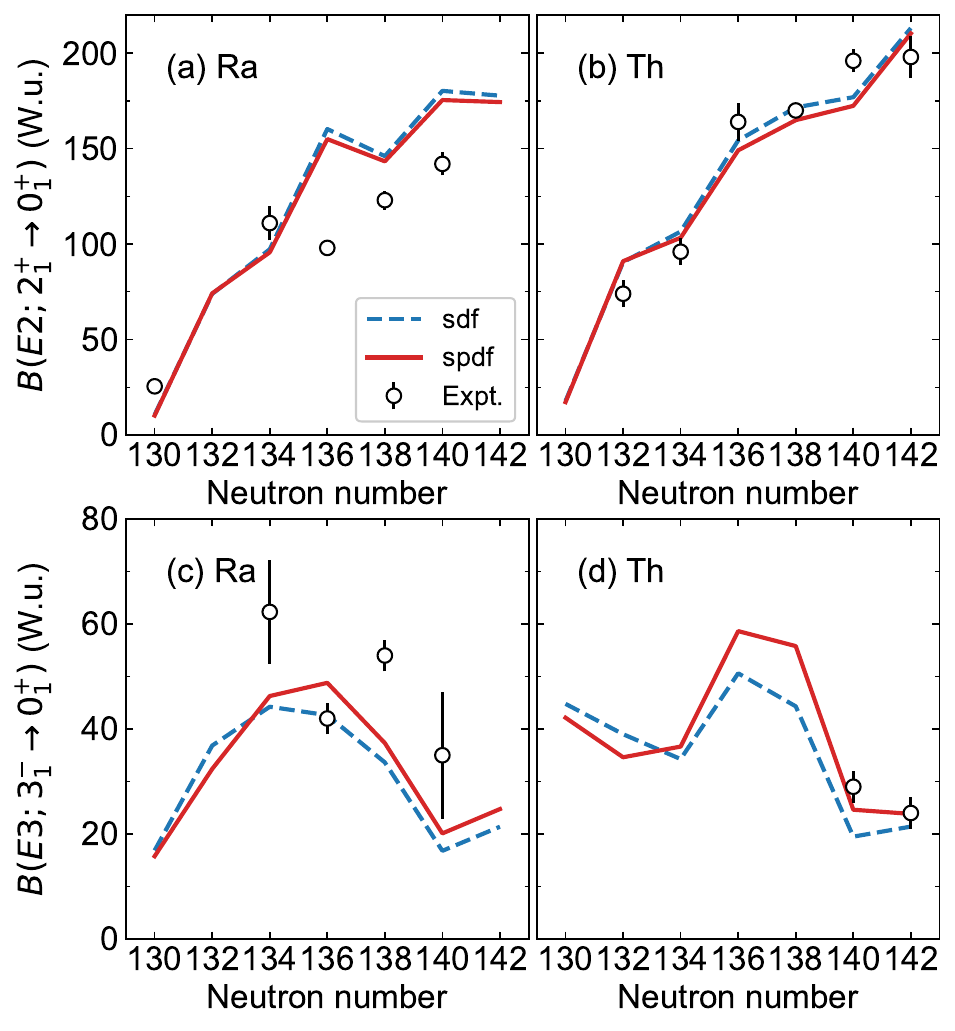}\\
\caption{Calculated $B(E2)$ and $B(E3)$ values in
Weisskopf units (W.u.) for
$^{218-230}$Ra and $^{220-232}$Th in the
$sdf$-IBM and $spdf$-IBM, and the corresponding experimental
data \cite{data,gaffney2013,butler2020a,chishti2020}.}
\label{fig:e23}
\end{center}
\end{figure}

%
%
\begin{figure}[htb!]
\begin{center}
\includegraphics[width=\linewidth]{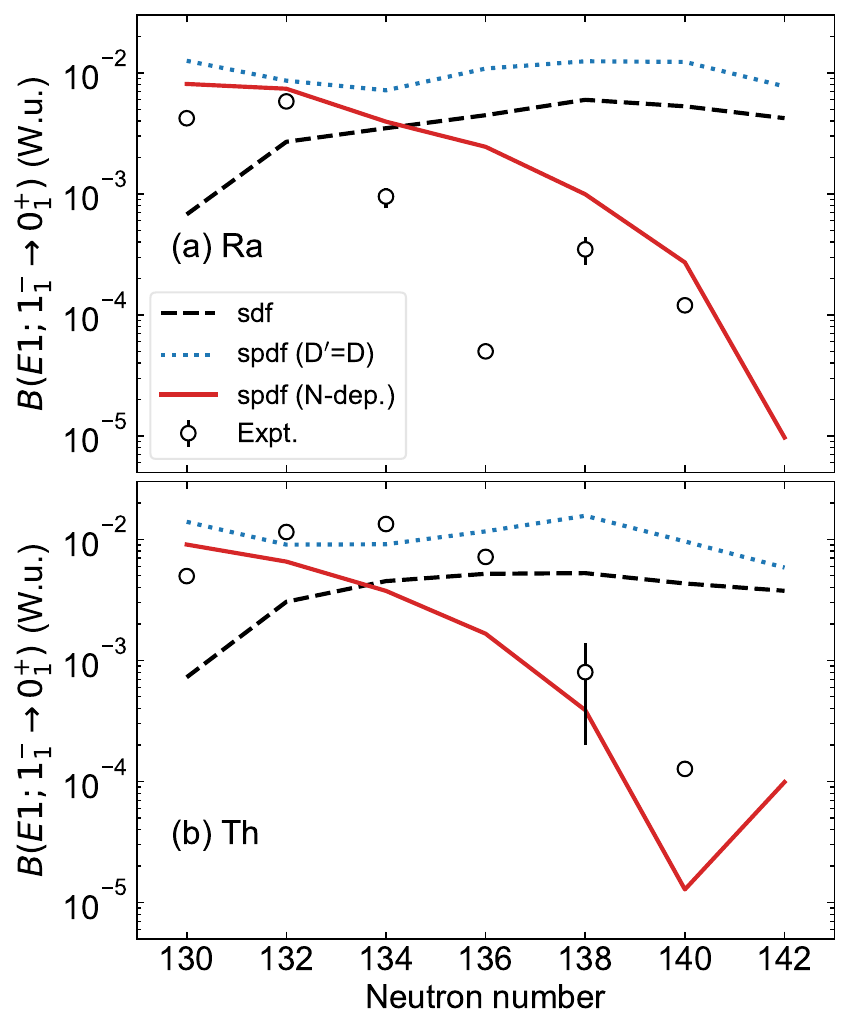}\\
\caption{Calculated $B(E1;1^-_1 \to 0^+_1)$ values in W.u.
for $^{218-230}$Ra and $^{220-232}$Th within the mapped
$sdf$-IBM and $spdf$-IBM.
Two $spdf$-IBM results denoted D$'=$ D and $N$-dep.
refer, respectively, to the calculations with the $E1$ operator
defined by \eqref{eq:e1} and that defined as $\hat D'=\hat D$
with the constant effective charge $e_1=0.008$ $e$b$^{1/2}$.
Experimental data are adopted from \cite{data,wollersheim93,gaffney2013,butler2020a,chishti2020}.
Note that the experimental data for $^{224}$Ra
and $^{228}$Ra shown in (a) are upper and lower
limits, respectively.}
\label{fig:e1}
\end{center}
\end{figure}

%
%
\begin{figure}[htb!]
\begin{center}
\includegraphics[width=\linewidth]{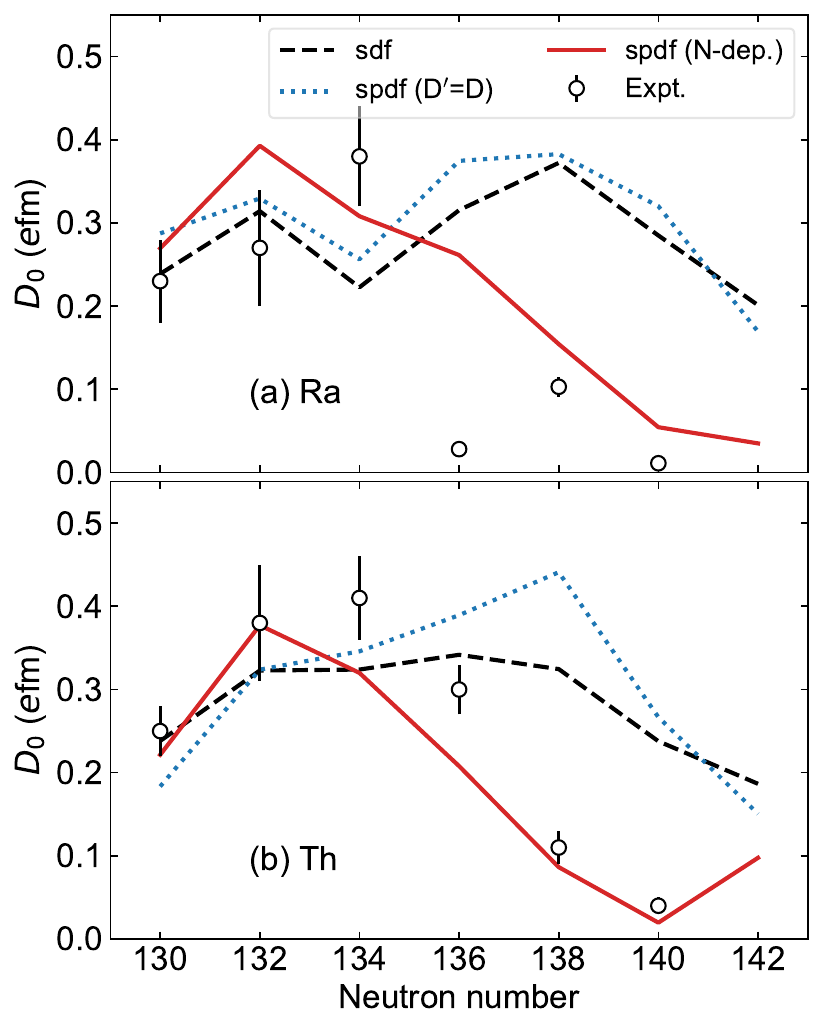}\\
\caption{Calculated $E1$ moment $D_0$ in $e$fm
for $^{218-230}$Ra and $^{220-232}$Th in the $sdf$-IBM,
and in the $spdf$-IBM using the $E1$ operator
with $\hat D'=\hat D$ and that which depends on the
neutron number $N$ defined in \eqref{eq:e1}.
Experimental data are taken from Ref.~\cite{butler1991}.}
\label{fig:d0}
\end{center}
\end{figure}

\subsection{Transition properties}

Figure~\ref{fig:e23} exhibits the $B(E2;2^+_1 \to 0^+_1)$
and $B(E3;3^-_1 \to 0^+_1)$ values in Weisskopf units (W.u.)
calculated by using the corresponding
transition operators in \eqref{eq:tel} with the
boson charges determined by the formulas of Eqs.~\eqref{eq:ech-q}
and \eqref{eq:ech-o}.
The $B(E2)$ values only increase with $N$ as the
quadrupole collectivity develops.
The calculated $B(E2)$ values are of the same order of
magnitudes as the experimental values
\cite{data,gaffney2013,butler2020a,chishti2020},
except for $^{224}$Ra ($N=136$), for which the
mapped IBM overestimates the data approximately
by a factor of 1.5.
In addition, the $B(E2)$ values obtained from
the mapped $spdf$-IBM do not differ quantitatively
from those from the $sdf$-IBM.
The calculated $B(E3;3^-_1 \to 0^+_1)$ values,
given in Figs.~\ref{fig:e23}(c) and \ref{fig:e23}(d),
show an inverse parabola with the maximal value near
$N=134$ (Ra) and $N=136$ (Th).
This behavior of the $B(E3)$ values confirms
that the octupole collectivity is most enhanced near $N=136$,
and correlates with the finding that the calculated negative-parity
energy levels are lowest in energy near those neutron
numbers.
The $B(E3)$ values obtained from the mapped $spdf$-IBM
are slightly smaller than those in the $sdf$-IBM for $N=130$ and 132,
whereas in octupole deformed region with $N\geqslant134$
more enhanced $E3$ transitions are suggested by the
$spdf$-IBM.

Figure~\ref{fig:e1} depicts the predicted $B(E1;1^-_1 \to 0^+_1)$
values in W.u., for $^{218-230}$Ra and $^{220-232}$Th.
The $sdf$-IBM calculation
with the $E1$ operator of the form of \eqref{eq:e1-sdf}
indicates a slight increase from $N=130-134$,
but suggests only a gradual variation
for $N\geqslant 136$, which contradicts the data. 
The $B(E1;1^-_1 \to 0^+_1)$ values in the $spdf$-IBM
obtained by using the $E1$
operator in which for $\hat D'$ the same dipole operator
$\hat D$ \eqref{eq:d} as in the $spdf$-boson
Hamiltonian \eqref{eq:bh} is used (denoted D$'=$ D
in the figure) do not exhibit any significant
variation with $N$.
The $B(E1)$ values obtained from the $spdf$-IBM with
the $E1$ operator defined in \eqref{eq:e1} ($N$-dep.)
show an overall decrease by orders of magnitudes
as functions of $N$,
consistently with the experimental systematic.
A significant deviation from the experimental data in
either choice of the $E1$ transition operator
in the $spdf$-IBM is found for $^{224}$Ra
for which the present calculations provide the
$B(E1;1^-_1 \to 0^+_1)$ value that is by orders
of magnitude larger than the experimentally estimated
upper limit \cite{gaffney2013}.

Figure~\ref{fig:d0} shows the predicted intrinsic
dipole moment of the studied Ra and Th nuclei
within the $sdf$-IBM and
$spdf$-IBM using the $E1$ operator with $\hat D'=\hat D$
and that which depends on the neutron number $N$ \eqref{eq:e1}.
The experimental dipole moments \cite{butler1991} included in
the figure were those obtained as averages of transition moments
of some high-spin states with $I\geqslant6$.
As in Ref.~\cite{otsuka1988},
the $D_0$ moments are here obtained from the
calculated $B(E1;6^+_1 \to 5^-_1)$ transition rates,
and using the formula
\begin{eqnarray}
 B(E1;I \to I') = \frac{3}{4\pi} D_0^2 (I1K0|I'K') \; ,
\end{eqnarray}
where $(I1K0|I'K')$ denotes a Clebsch-Gordan coefficient,
and the quantum numbers $K=K'=0$ for the initial ($I$)
and final ($I'$) states, respectively.
It is seen in the figure that the calculated
$D_0$ moments in the $spdf$-IBM with the
$E1$ operator of \eqref{eq:e1} exhibit systematic and
absolute values that are consistent with
the experimental data, except for the $^{224}$Ra nucleus.
The calculation in the $sdf$-IBM, and the one
in the $spdf$-IBM with the $E1$ operator
corresponding to the same dipole
operator as in the Hamiltonian do not appear to be
able to reproduce the correct observed behaviors
of this quantity for deformed region, $N\geqslant138$.

%
%
\begin{figure*}[htb!]
\begin{center}
\includegraphics[width=.49\linewidth]{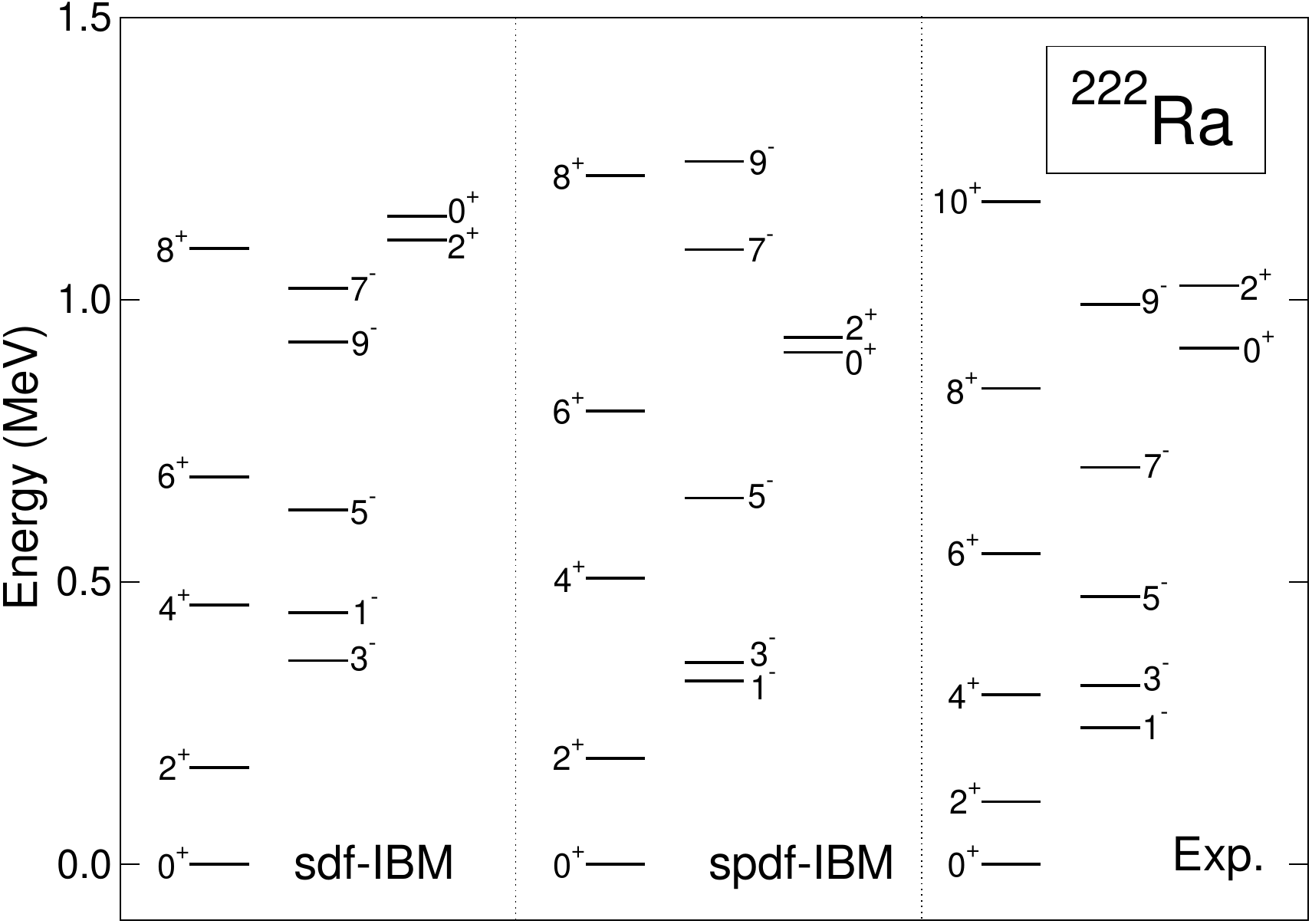}
\includegraphics[width=.49\linewidth]{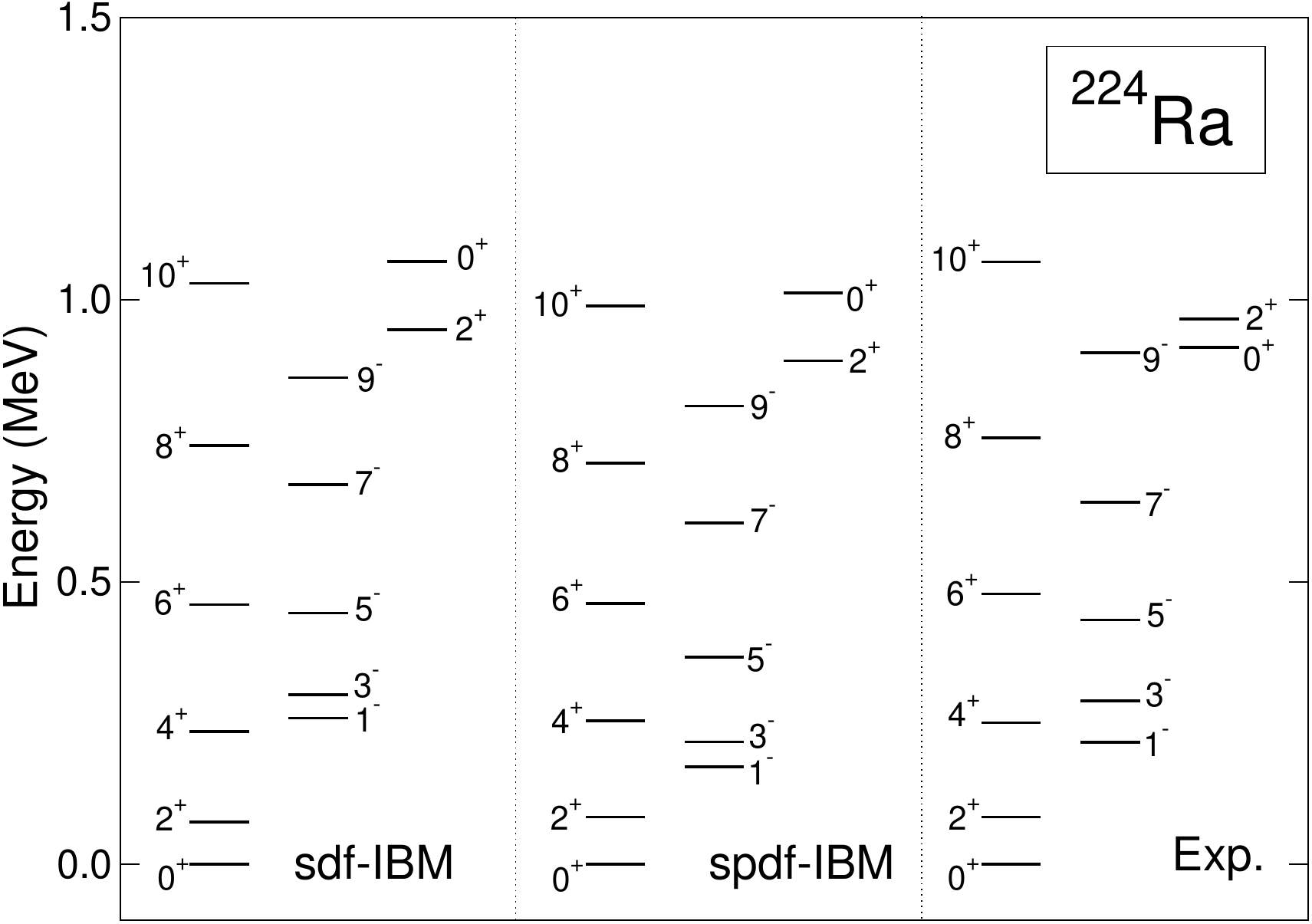}\\
\includegraphics[width=.49\linewidth]{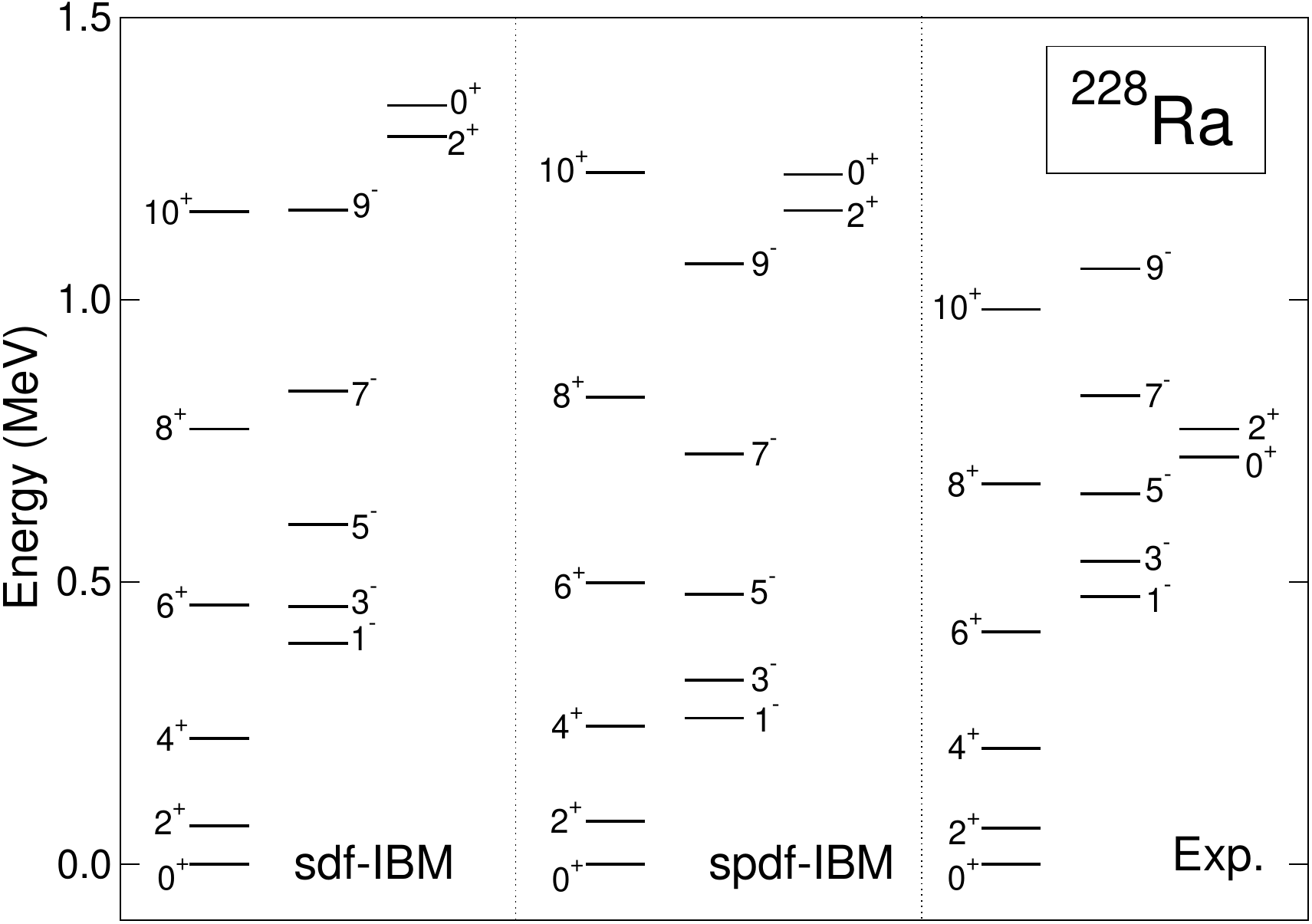}
\includegraphics[width=.49\linewidth]{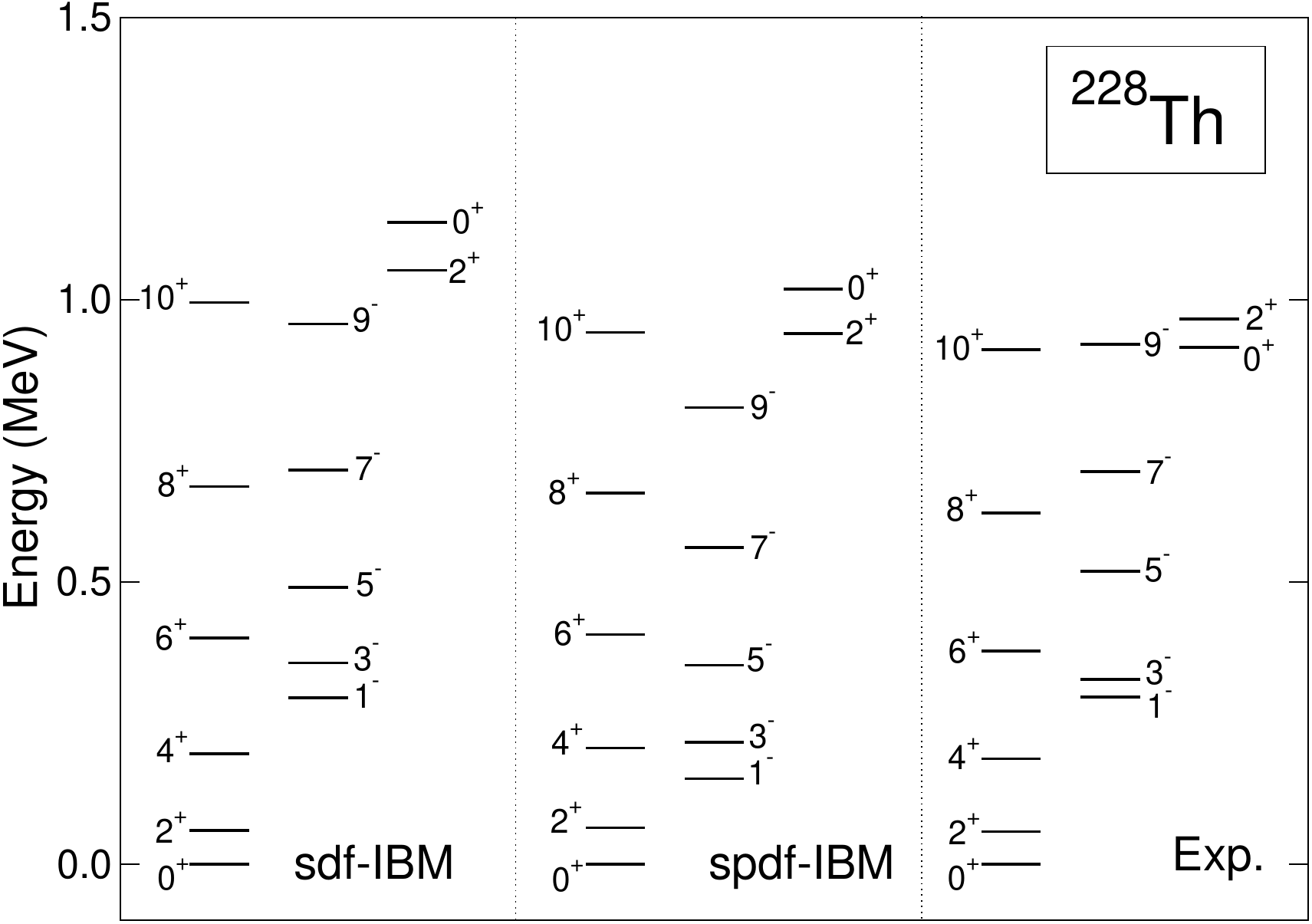}\\
\caption{Low-energy spectra for $^{222}$Ra, $^{224}$Ra, $^{228}$Ra,
and $^{228}$Th calculated with the
mapped $sdf$-IBM and $spdf$-IBM, and the corresponding
experimental data \cite{data,gaffney2013,chishti2020}.}
\label{fig:spec}
\end{center}
\end{figure*}

\subsection{Detailed energy spectra}

Figure~\ref{fig:spec} displays the calculated
low-energy spectra of both parities
for the isotopes $^{222}$Ra, $^{224}$Ra, $^{228}$Ra,
and $^{228}$Th within the mapped $sdf$-IBM and $spdf$-IBM.
These nuclei are here chosen with a renewed interest from
an experimental point of view, since updated
data have recently been provided
\cite{gaffney2013,butler2020a,chishti2020,data}.
Comparisons of the theoretical and experimental $E2$, $E3$,
and $E1$ transition properties for these nuclei
are given in Tables~\ref{tab:ra222}--\ref{tab:th228}.

%
%
\begin{table}[hb!]
\caption{\label{tab:ra222}
$B(E\lambda)$ transition strengths in W.u.
for $^{222}$Ra calculated within the $sdf$-IBM and $spdf$-IBM.
The corresponding experimental data are taken from NNDC \cite{data}.}
\begin{center}
 \begin{ruledtabular}
\begin{tabular}{cccc}
\textrm{$B(E\lambda;I^{\pi}_i \to I^{\pi}_f)$} &
\textrm{$sdf$-IBM} &
\textrm{$spdf$-IBM} &
\textrm{Expt.} \\
\hline
$ B(E2; {2}^{+}_{1} \to {0}^{+}_{1})$ & $97$ & $96$ & $112.8^{+9.6}_{-8.2}$ \\ 
$ B(E2; {4}^{+}_{1} \to {2}^{+}_{1})$ & $134$ & $131$ & $123\pm14$ \\ 
$ B(E2; {6}^{+}_{1} \to {4}^{+}_{1})$ & $134$ & $121$ & $135^{+24}_{-21}$ \\ 
$ B(E2; {8}^{+}_{1} \to {6}^{+}_{1})$ & $112$ & $99$ & $119^{+36}_{-28}$ \\ 
$ B(E2; {10}^{+}_{1} \to {8}^{+}_{1})$ & $82$ & $92$ & $139^{+76}_{-48}$ \\ 
$ B(E2; {3}^{-}_{1} \to {1}^{-}_{1})$ & $96$ & $103$ & $98^{+49}_{-40}$ \\ 
$ B(E2; {5}^{-}_{1} \to {3}^{-}_{1})$ & $116$ & $122$ & $109^{+30}_{-26}$ \\ 
$ B(E2; {7}^{-}_{1} \to {5}^{-}_{1})$ & $117$ & $121$ & $95^{+74}_{-38}$ \\ 
$ B(E2; {9}^{-}_{1} \to {7}^{-}_{1})$ & $45$ & $65$ & $2.7^{+1.0}_{-0.7}\times10^{2}$ \\ 
$ B(E1; {1}^{-}_{1} \to {0}^{+}_{1})$ & $3.5\times10^{-3}$ & $3.9\times10^{-3}$ & $9.5^{+1.9}_{-1.7}\times10^{-4}$ \\ 
$ B(E1; {1}^{-}_{1} \to {2}^{+}_{1})$ & $3.3\times10^{-3}$ & $6.3\times10^{-3}$ & $1.86^{+0.40}_{-0.34}\times10^{-3}$ \\ 
$ B(E1; {3}^{-}_{1} \to {2}^{+}_{1})$ & $4.9\times10^{-3}$ & $4.6\times10^{-3}$ & $4.1^{+1.7}_{-1.4}\times10^{-3}$ \\ 
$ B(E1; {5}^{-}_{1} \to {4}^{+}_{1})$ & $5.6\times10^{-3}$ & $4.5\times10^{-3}$ & $1.23^{+0.60}_{-0.24}\times10^{-3}$ \\ 
$ B(E1; {6}^{+}_{1} \to {5}^{-}_{1})$ & $2.3\times10^{-3}$ & $4.4\times10^{-3}$ & $2.11^{+0.44}_{-0.40}\times10^{-3}$ \\ 
$ B(E1; {7}^{-}_{1} \to {6}^{+}_{1})$ & $6.5\times10^{-3}$ & $4.8\times10^{-3}$ & $2.4^{+1.2}_{-2.1}\times10^{-3}$ \\ 
$ B(E1; {8}^{+}_{1} \to {7}^{-}_{1})$ & $4.8\times10^{-3}$ & $5.2\times10^{-3}$ & $2.65^{+0.75}_{-0.66}\times10^{-3}$ \\ 
$ B(E1; {9}^{-}_{1} \to {8}^{+}_{1})$ & $6.3\times10^{-3}$ & $4.9\times10^{-3}$ & $4.9^{+1.8}_{-1.4}\times10^{-3}$ \\ 
$ B(E1; {10}^{+}_{1} \to {9}^{-}_{1})$ & $4.5\times10^{-3}$ & $4.3\times10^{-3}$ & $3.2^{+1.6}_{-1.1}\times10^{-3}$ \\
\end{tabular}
 \end{ruledtabular}
\end{center}
\end{table}

%
%
\begin{table}[hb!]
\caption{\label{tab:ra222-rme}
Calculated and experimental \cite{butler2020a} reduced
$E2$ and $E3$ matrix elements in $e$b$^{\lambda/2}$ units for $^{222}$Ra.
}
\begin{center}
 \begin{ruledtabular}
\begin{tabular}{cccc}
\textrm{$\braket{I\|E\lambda\|I'}$} &
\textrm{$sdf$-IBM} &
\textrm{$spdf$-IBM} &
\textrm{Expt.} \\
\hline
$\braket{{2}^{+}_{1}||E2||{2}^{+}_{1}}$ & $-2.4$ & $-2.3$ & $-1.3\pm0.5$ \\ 
$\braket{{4}^{+}_{1}||E2||{2}^{+}_{1}}$ & $-3.1$ & $-3.1$ & $2.98\pm0.15$ \\ 
$\braket{{6}^{+}_{1}||E2||{4}^{+}_{1}}$ & $3.7$ & $3.5$ & $3.57\pm0.18$ \\ 
$\braket{{8}^{+}_{1}||E2||{6}^{+}_{1}}$ & $3.9$ & $3.7$ & $4.15\pm0.23$ \\ 
$\braket{{3}^{-}_{1}||E2||{1}^{-}_{1}}$ & $-2.3$ & $2.4$ & $2.35\pm0.22$ \\ 
$\braket{{5}^{-}_{1}||E2||{3}^{-}_{1}}$ & $3.2$ & $-3.3$ & $3.1\pm0.4$ \\ 
$\braket{{7}^{-}_{1}||E2||{5}^{-}_{1}}$ & $-3.7$ & $-3.8$ & $4.4\pm0.4$ \\ 
$\braket{{9}^{-}_{1}||E2||{7}^{-}_{1}}$ & $-2.6$ & $3.1$ & $6.0\pm1.0$ \\ 
$\braket{{1}^{-}_{1}||E3||{2}^{+}_{1}}$ & $0.9$ & $-1.0$ & $0.85\pm0.24$ \\ 
$\braket{{3}^{-}_{1}||E3||{0}^{+}_{1}}$ & $-1.0$ & $-1.0$ & $1.13\pm0.09$ \\ 
$\braket{{3}^{-}_{1}||E3||{2}^{+}_{1}}$ & $1.0$ & $1.1$ & $-0.9\pm0.5$ \\ 
$\braket{{3}^{-}_{1}||E3||{4}^{+}_{1}}$ & $0.9$ & $1.0$ & $2.6^{+0.6}_{-0.9}$ \\ 
$\braket{{4}^{+}_{1}||E3||{1}^{-}_{1}}$ & $-0.8$ & $0.9$ & $-2.1\pm0.5$ \\ 
$\braket{{5}^{-}_{1}||E3||{2}^{+}_{1}}$ & $-1.4$ & $1.4$ & $1.79\pm0.20$ \\ 
$\braket{{5}^{-}_{1}||E3||{4}^{+}_{1}}$ & $-1.2$ & $1.3$ & $-1.7\pm1.0$ \\ 
$\braket{{7}^{-}_{1}||E3||{4}^{+}_{1}}$ & $-1.6$ & $1.7$ & $3.3^{+0.3}_{-0.5}$ \\\end{tabular}
 \end{ruledtabular}
\end{center}
\end{table}

In Fig.~\ref{fig:spec} the calculated ground-state band
in both the $sdf$- and $spdf$-boson models for $^{222}$Ra
is rather stretched in comparison with experiment.
As mentioned above, the $K=0^-$
band in the $sdf$-IBM exhibits inversions of levels $1^-$ and $3^-$,
and $7^-$ and $9^-$.
The wrong order of levels is lifted in the $spdf$-IBM
energy spectra. Neither of the two mapped boson models,
however, produces the $K=0^-$ band with moment of inertia
consistent with that of the observed $K=0^-$ band.
The irregularity in the predicted $K=0^-$ bands
could be attributed to the fact that significant amounts
of $f$-boson configurations enter the wave functions
of high-spin states with spin $I\geqslant7$
[see Fig.~\ref{fig:wf}(c)].
The predicted ground-state bands in $sdf$ and $spdf$
frameworks also exhibit some irregularity,
i.e., the $6^+$ energy level is rather close to the $4^+$ level.
The $f$-boson contributions appear to play a role
also in these positive-parity states.
The IBM mapping calculations yield the non-yrast $0^+_2$
and $2^+_2$ levels with energies consistent with
the experimental values \cite{data}.
From Fig.~\ref{fig:spec} one can see that the mapped
$spdf$-IBM provides an improved description of
these levels over the $sdf$-IBM.
In the $sdf$-IBM, for both the $0^+_2$
and $2^+_2$ states the expectation value
$\braket{\hat n_f}\approx2$, indicating the double-octupole
phonon nature as suggested in the earlier mapped
$sdf$-IBM calculation based on the relativistic
EDF \cite{nomura2014} and in some
experimental studies, e.g., Ref.~\cite{spieker2013}.
The $spdf$-IBM wave functions of these states are
characterized by dominant $f$-boson contributions
with the expectation value $\braket{\hat n_f}\approx1.3$
and the $p$-boson contributions with
$\braket{\hat n_p}\approx0.6$.

As shown in Table~\ref{tab:ra222}, 
by the inclusion of $p$ bosons
$B(E2)$ values for the in-band transitions
in the ground-state band for $^{222}$Ra
are generally reduced, while those in the negative-parity
band are enhanced.
The $B(E1)$ rates are,
in general, of the same order of magnitude
as the experimental values \cite{data}.
In Table~\ref{tab:ra222-rme}
the calculated $E2$ and $E3$ reduced matrix elements
are compared with the recent experimental data \cite{butler2020a}.
Up to sign most of the predicted matrix elements
are consistent with the data.
The $spdf$-IBM results for the $E3$ matrix elements
are overall larger in magnitude than those in the $sdf$-IBM.

%
%
\begin{table}[hb!]
\caption{\label{tab:ra224}
Same as the caption to Table~\ref{tab:ra222}, but for $^{224}$Ra.
Experimental data are from Refs.~\cite{gaffney2013,data}.}
\begin{center}
 \begin{ruledtabular}
\begin{tabular}{cccc}
\textrm{$B(E\lambda;I^{\pi}_i \to I^{\pi}_f)$} &
\textrm{$sdf$-IBM} &
\textrm{$spdf$-IBM} &
\textrm{Expt.} \\
\hline
$ B(E2; {2}^{+}_{1} \to {0}^{+}_{1})$ & $160$ & $155$ & $98\pm3$ \\ 
$ B(E2; {2}^{+}_{2} \to {0}^{+}_{1})$ & $0.8$ & $1.2$ & $1.3\pm0.5$ \\ 
$ B(E2; {4}^{+}_{1} \to {2}^{+}_{1})$ & $221$ & $210$ & $137\pm5$ \\ 
$ B(E2; {6}^{+}_{1} \to {4}^{+}_{1})$ & $222$ & $195$ & $156\pm12$ \\ 
$ B(E2; {8}^{+}_{1} \to {6}^{+}_{1})$ & $201$ & $184$ & $180\pm60$ \\ 
$ B(E2; {3}^{-}_{1} \to {1}^{-}_{1})$ & $168$ & $173$ & $93\pm9$ \\ 
$ B(E2; {5}^{-}_{1} \to {3}^{-}_{1})$ & $199$ & $204$ & $190\pm60$ \\ 
$ B(E3; {1}^{-}_{1} \to {2}^{+}_{1})$ & $106$ & $128$ & $210\pm40$ \\ 
$ B(E3; {3}^{-}_{1} \to {0}^{+}_{1})$ & $43$ & $49$ & $42\pm3$ \\ 
$ B(E3; {3}^{-}_{1} \to {2}^{+}_{1})$ & $52$ & $62$ & $<600$ \\ 
$ B(E3; {5}^{-}_{1} \to {2}^{+}_{1})$ & $61$ & $70$ & $61\pm17$ \\ 
$ B(E1; {1}^{-}_{1} \to {0}^{+}_{1})$ & $4.5\times10^{-3}$ & $2.4\times10^{-3}$ & $<5\times10^{-5}$ \\ 
$ B(E1; {1}^{-}_{1} \to {2}^{+}_{1})$ & $5.8\times10^{-3}$ & $4.0\times10^{-3}$ & $<1.3\times10^{-4}$ \\ 
$ B(E1; {3}^{-}_{1} \to {2}^{+}_{1})$ & $6.4\times10^{-3}$ & $3.1\times10^{-3}$ & $3.9^{+1.7}_{-1.4}\times10^{-5}$ \\ 
$ B(E1; {5}^{-}_{1} \to {4}^{+}_{1})$ & $7.3\times10^{-3}$ & $3.2\times10^{-3}$ & $4^{+3}_{-2}\times10^{-5}$ \\ 
$ B(E1; {7}^{-}_{1} \to {6}^{+}_{1})$ & $8.4\times10^{-3}$ & $3.3\times10^{-3}$ & $<3\times10^{-4}$ \\
\end{tabular}
 \end{ruledtabular}
\end{center}
\end{table}

The nucleus $^{224}$Ra was experimentally revealed to
be empirical evidence for the permanent octupole deformation
\cite{gaffney2013}.
The $(\beta_2,\beta_3)$-PES of this nucleus exhibits
a distinct minimum with $\beta_3\neq0$, and an approximate
alternating parity band is suggested experimentally.
It is seen from Fig.~\ref{fig:spec} that
the mapped IBM calculations in both the $sdf$ and $spdf$
versions yield low-energy spectra of states with both
parities that are consistent with the experimental data.
An improvement over the $sdf$-IBM could be that
the $1^-$ bandhead of the $K=0^-$ band in the $spdf$-IBM
is found below the $4^+$ level, which is indeed observed
experimentally.
The non-yrast $0^+_2$ and $2^+_2$ levels are lowered in
energy in the $spdf$-IBM, but the order of these levels
is at variance with experiment.
However, the calculated $2^+_2$ state may here be of
different character, e.g.,
bandhead of the $\gamma$-vibrational band, from the $0^+_2$ state,
and probably should not be considered
a member of the band built on the $0^+_2$ state.
Both of these states are of double-octupole phonon nature
in the present $sdf$-IBM, while $f$ and $p$ boson
configurations make appreciable contributions
to the wave functions in the $spdf$-IBM,
since the expectation values
$\braket{\hat n_f}\approx1.5$ and
$\braket{\hat n_p}\approx0.3$.
As shown in Table~\ref{tab:ra224},
the calculated $B(E2)$ values generally overestimate
the experimental values, while the $B(E3)$ values
obtained in either of the two boson models are
more or less within error bars of the experimental data.
As is well known, the $B(E1)$ transition rates for
$^{224}$Ra are anomalously small as compared with
the neighboring isotopes $^{222}$Ra and $^{226}$Ra
[see also Fig.~\ref{fig:e1}(a)].
The present IBM mapping calculations,
even that which includes the $p$-boson degrees of freedom,
fail to reproduce this observed systematic.

%
%
\begin{table}[hb!]
\caption{\label{tab:ra228}
Same as the caption to Table~\ref{tab:ra222}, but for $^{228}$Ra.
Experimental data are from Ref.~\cite{data}.}
\begin{center}
 \begin{ruledtabular}
\begin{tabular}{cccc}
\textrm{$B(E\lambda;I^{\pi}_i \to I^{\pi}_f)$} &
\textrm{$sdf$-IBM} &
\textrm{$spdf$-IBM} &
\textrm{Expt.} \\
\hline
$ B(E2; {2}^{+}_{1} \to {0}^{+}_{1})$ & $180$ & $175$ & $142\pm6$ \\ 
$ B(E2; {4}^{+}_{1} \to {2}^{+}_{1})$ & $253$ & $246$ & $207\pm4$ \\ 
$ B(E1; {1}^{-}_{1} \to {0}^{+}_{1})$ & $5.3\times10^{-3}$ & $2.7\times10^{-4}$ & $\geq1.2\times10^{-4}$ \\ 
$ B(E1; {1}^{-}_{1} \to {2}^{+}_{1})$ & $7.5\times10^{-3}$ & $3.4\times10^{-4}$ & $\geq1.5\times10^{-4}$ \\ 
$ B(E1; {3}^{-}_{1} \to {2}^{+}_{1})$ & $7.6\times10^{-3}$ & $3.7\times10^{-4}$ & $\geq2.2\times10^{-4}$ \\ 
$ B(E1; {3}^{-}_{1} \to {4}^{+}_{1})$ & $5.1\times10^{-3}$ & $2.0\times10^{-4}$ & $\geq1.5\times10^{-4}$ \\
\end{tabular}
 \end{ruledtabular}
\end{center}
\end{table}

%
%
\begin{table}[hb!]
\caption{\label{tab:ra228-rme}
Same as the caption to Table~\ref{tab:ra222-rme},
but for $^{228}$Ra.
}
\begin{center}
 \begin{ruledtabular}
\begin{tabular}{cccc}
\textrm{$\braket{I\|E\lambda\|I'}$} &
\textrm{$sdf$-IBM} &
\textrm{$spdf$-IBM} &
\textrm{Expt.} \\
\hline
$\braket{{2}^{+}_{1}||E2||{2}^{+}_{1}}$ & $-3.3$ & $-3.2$ & $-0.3\pm1.7$ \\ 
$\braket{{4}^{+}_{1}||E2||{2}^{+}_{1}}$ & $-4.3$ & $-4.3$ & $3.87\pm0.19$ \\ 
$\braket{{6}^{+}_{1}||E2||{4}^{+}_{1}}$ & $5.4$ & $5.3$ & $5.11\pm0.26$ \\ 
$\braket{{8}^{+}_{1}||E2||{6}^{+}_{1}}$ & $6.1$ & $-6.0$ & $5.89\pm0.29$ \\ 
$\braket{{10}^{+}_{1}||E2||{8}^{+}_{1}}$ & $-6.6$ & $6.4$ & $7.5\pm0.4$ \\ 
$\braket{{3}^{-}_{1}||E2||{1}^{-}_{1}}$ & $3.4$ & $3.4$ & $3.8\pm0.5$ \\ 
$\braket{{5}^{-}_{1}||E2||{3}^{-}_{1}}$ & $4.6$ & $-4.6$ & $3.9^{+0.4}_{-0.8}$ \\ 
$\braket{{7}^{-}_{1}||E2||{5}^{-}_{1}}$ & $5.5$ & $5.5$ & $4.0\pm0.9$ \\ 
$\braket{{9}^{-}_{1}||E2||{7}^{-}_{1}}$ & $6.1$ & $-6.1$ & $5.9\pm1.0$ \\ 
$\braket{{1}^{-}_{1}||E3||{2}^{+}_{1}}$ & $0.6$ & $-0.7$ & $1.36\pm0.23$ \\ 
$\braket{{3}^{-}_{1}||E3||{0}^{+}_{1}}$ & $-0.6$ & $0.7$ & $0.87\pm0.15$ \\ 
$\braket{{3}^{-}_{1}||E3||{2}^{+}_{1}}$ & $-0.7$ & $0.7$ & $-0.06^{+0.23}_{-0.16}$ \\ 
$\braket{{4}^{+}_{1}||E3||{1}^{-}_{1}}$ & $-0.6$ & $0.7$ & $0.4^{+0.7}_{-1.1}$ \\ 
$\braket{{5}^{-}_{1}||E3||{2}^{+}_{1}}$ & $0.9$ & $1.0$ & $1.71\pm0.23$ \\
\end{tabular}
 \end{ruledtabular}
\end{center}
\end{table}

The $(\beta_2,\beta_3)$-PES of the $^{228}$Ra nucleus
is shown to be particularly soft along the $\beta_3$
deformation. The observed energy spectra show features that
the $K=0^-$ yrast band decouples from the
ground-state $K=0^+$ band, forming separate
rotational bands \cite{data}.
This feature is reproduced in the $sdf$-IBM, since
as shown in Fig.~\ref{fig:spec} the $1^-$ bandhead of
the $K=0^-$ band appears close to the $6^+$ level.
The calculated $K=0^-$ band in the $spdf$-IBM is, however,
much lower than the experimental counterpart.
This suggests that the contributions of the negative-parity
bosons to the low-lying $K=0^-$ band for $^{228}$Ra
are rather overestimated in the $spdf$-IBM.
The $0^+_2$ and $2^+_2$ states in the $spdf$-IBM
are lower in energy than those in the $sdf$-IBM, but
are considerably higher than the experimental values.
These non-yrast states are predicted to be
formed by the coupling of approximately two $f$ bosons in
the $sdf$ model, and are in the $spdf$ model
comprised of large fractions of the $f$-boson
($\braket{\hat n_f}\approx1.7$) and minor contributions
from the $p$-boson components
($\braket{\hat n_p}\approx0.2$).

In Table~\ref{tab:ra228}
the mapped $spdf$-IBM gives smaller $B(E2)$ values
than the $sdf$-IBM for $^{228}$Ra.
The calculated $B(E1)$ values in the $spdf$-IBM
are by an order magnitude smaller than those in the
$sdf$-IBM, and are of the same order of magnitude
as the experimental data \cite{data}.
The $E2$ and $E3$ reduced matrix elements
are listed in Table~\ref{tab:ra228-rme}, in which
the calculated results are compared with the
recent data \cite{butler2020a}.
Apart from the sign these calculated matrix elements are
mostly consistent with the experimental values.

%
%
\begin{table}[hb!]
\caption{\label{tab:th228}
Same as the caption to Table~\ref{tab:ra222}, but for $^{228}$Th.
Experimental data are from Ref.~\cite{chishti2020}.}
\begin{center}
 \begin{ruledtabular}
\begin{tabular}{cccc}
\textrm{$B(E\lambda;I^{\pi}_i \to I^{\pi}_f)$} &
\textrm{$sdf$-IBM} &
\textrm{$spdf$-IBM} &
\textrm{Expt.} \\
\hline
$ B(E2; {2}^{+}_{1} \to {0}^{+}_{1})$ & $172$ & $165$ & $170\pm3$ \\ 
$ B(E2; {4}^{+}_{1} \to {2}^{+}_{1})$ & $240$ & $229$ & $224\pm12$ \\ 
$ B(E2; {2}^{+}_{2} \to {0}^{+}_{1})$ & $0.7$ & $1.4$ & $1.1\pm0.6$ \\ 
$ B(E2; {2}^{+}_{2} \to {2}^{+}_{1})$ & $0.7$ & $0.7$ & $2.5\pm1.4$ \\ 
$ B(E1; {1}^{-}_{1} \to {0}^{+}_{1})$ & $5.3\times10^{-3}$ & $3.9\times10^{-4}$ & $(8\pm6)\times10^{-4}$ \\ 
$ B(E1; {1}^{-}_{1} \to {2}^{+}_{1})$ & $8.0\times10^{-3}$ & $5.9\times10^{-4}$ & $\geq1.4\times10^{-3}$ \\ 
$ B(E1; {3}^{-}_{1} \to {2}^{+}_{1})$ & $7.6\times10^{-3}$ & $5.1\times10^{-4}$ & $(3.8\pm0.7)\times10^{-4}$ \\
\end{tabular}
 \end{ruledtabular}
\end{center}
\end{table}

A measurement of the $E1$ moment was recently
made for $^{228}$Th \cite{chishti2020}, producing some data
to be compared with the present model results.
The corresponding $(\beta_2,\beta_3)$-PES is rather
soft in $\beta_3$ deformation, but exhibits
a nonzero $\beta_3$ minimum, suggesting presence
of strong octupole correlations.
The observed energy spectra, shown in Fig.~\ref{fig:spec},
suggest the $K=0^-$ bandhead to be in between the $4^+$ and $6^+$
levels, but do not appear to form
an alternating-parity band.
The $spdf$-IBM gives the $K=0^-$ band that is much lower than
the experimental one.
The $sdf$-IBM calculation, however, reproduces
the observed yrast bands of both parities rather well,
and it appears that the $sdf$ framework may be sufficient
to account for the $^{228}$Th spectra.
The $0^+_2$ and $2^+_2$ non-yrast levels are more
accurately reproduced in the $spdf$-IBM.
Also, as shown in Table~\ref{tab:th228},
the calculated $B(E2)$ values in the $spdf$-IBM
reproduce the experimental values more accurately
than the $sdf$-IBM.
A significant improvement over the $sdf$-boson model
is that the orders of magnitude of the calculated $B(E1)$
transition rates specifically for the
$1^-_1 \to 0^+_1$ and $3^-_1 \to 2^+_1$ transitions
are consistent with those of the experimental values.

\section{Conclusions\label{sec:summary}}

In the present work, dipole $p$ boson degrees of freedom
have been incorporated in the IBM mapping
for describing quadrupole-octupole collective states in heavy nuclei.
A procedure has been proposed that is to determine strength
parameters of the $spdf$-IBM Hamiltonian using as microscopic
inputs the HFB energy surfaces in terms of
the axially symmetric quadrupole-octupole $(\beta_2,\beta_3)$,
dipole-quadrupole $(\beta_1,\beta_2)$,
and dipole-octupole $(\beta_1,\beta_3)$ deformations.
With the assumptions that some of those parameters that appear
in the $spdf$-boson model are similar to those of
the $sdf$ model and that most of them stay constant with
nucleon number, it has been shown that one can obtain
a set of the model parameters with which the IBM
energy surfaces in the three deformation spaces
can be made similar as much as possible to
the HFB PESs,
and which are physically sound in the sense that
overall behaviors of the observed low-energy
positive-parity and negative-parity
yrast levels in the nuclei in the region of interest
are reasonably described.

An illustrative application of the mapped $spdf$-IBM
to axially deformed actinides $^{218-230}$Ra and $^{220-232}$Th
revealed that the influences of including $p$ bosons
in the mapping procedure are to lower the $K=0^-$ band,
and to improve significantly the description of the
observed behaviors of the $B(E1)$
values and intrinsic dipole moment with respect to the
$sdf$-IBM, if the nucleon-number-dependent $E1$ transition
operator that takes into account the underlying
shell structure is employed.
The effect of $p$ bosons in lowering the $1^-$ energy
level is most pronounced for those nuclei with $N\approx132$
corresponding to the transitional region from
the nearly spherical to static octupole deformed regimes.
The relevant $1^-$ wave functions of these nuclei
are accounted for by a large fraction of the $p$-boson
configurations, which dominate over those of $f$ bosons.
For higher-spin states and states in those nuclei in
the deformed region ($N\geqslant134$), however,
the $p$ boson contributions are minor, and most
of these states are dominated by the $f$-boson configurations.
In these heavier nuclei, in some cases the $sdf$-IBM
appears to give a sufficiently good description
of the excitation energies.
It was further shown that while there is essentially
no significant difference between the $sdf$-IBM and
$spdf$-IBM results for the $E2$ and $E3$
transition properties, the inclusion of $p$ bosons
reduces the $B(E1)$ values by orders of magnitude,
which are consistent with the experimental data, in particular,
for deformed heavy nuclei.
The mapped $spdf$-IBM provides
reasonable descriptions of the systematic
of the $1^-_1$ energy levels,
$B(E1;1^-_1 \to 0^+_1)$,
and $B(E3;3^-_1 \to 0^+_1)$ values,
which also compare those of
the beyond-mean-field
studies starting from the same
input, i.e., the constrained
HFB calculations
with the Gogny interactions,
e.g., in Ref.~\cite{robledo2013}.

Some difficulty with the model description was shown
to arise in calculating the $E1$ properties of the $^{224}$Ra
nucleus in particular.
The observed $B(E1;1^-_1 \to 0^+_1)$ value for this
nucleus is smaller than
those for the neighboring isotopes by
orders of magnitudes.
The present $spdf$-boson model is not able to
reproduce this local behavior,
whereas the previous beyond-mean-field calculation
\cite{robledo2013} accounted for it.
This indicates a limitation
of the present version of the mapping procedure.
An immediate remedy would be to devise
and test in a systematic manner
an $E1$ transition operator that is more realistic
than that exploited here from Ref.~\cite{otsuka1988}.
This presents an open problem for the IBM mapping to be addressed
in the future.

The $spdf$-IBM mapping procedure developed in the present study
opens up several possibilities of nuclear structure studies
related to octupole deformations.
It will be further applied to those nuclei in other
mass regions of interest
in which octupole correlations are supposed to be enhanced,
including the Sm-Gd nuclei in rare-earth regions,
and Ba-Xe nuclei in neutron-rich lanthanides.
These extended studies will allow for more accurate
predictions of the spectroscopic properties of low-energy
quadrupole-octupole collective states, which
attract considerable attention from experimental studies using
the RI beams and from the related theoretical investigations.
The $p$-boson effects could also be relevant in the
calculations of EDMs in octupole deformed odd-mass nuclei,
which are of fundamental importance in basic physics.
These will be interesting future studies, and related
results will be reported elsewhere.

\acknowledgements
This work has been supported by JSPS
KAKENHI Grant No. JP25K07293.

\bibliography{refs}

\end{document}